\documentclass[english,aps,prb,superscriptaddress,floatfix,reprint,twocolumn]{revtex4-1}
\usepackage[T1]{fontenc}
\usepackage[utf8]{inputenc}
\usepackage{mathrsfs}
\usepackage{mathtools}
\usepackage{amsmath}
\usepackage{amssymb}
\usepackage{cancel}
\usepackage{graphicx}

\makeatletter

\providecommand{\tabularnewline}{\\}

\usepackage[dvipsnames]{xcolor}
\usepackage{verbatim}
\usepackage{mathrsfs}
\usepackage{txfonts}

\providecommand{\tabularnewline}{\\}

\providecommand{\tabularnewline}{\\}

\providecommand{\tabularnewline}{\\}

\usepackage{physics}
\newcommand{\CC}{\mathrm C}
\newcommand{\JJ}{\mathrm J}
\newcommand{\EC}{E_\CC}
\newcommand{\EJ}{E_\JJ}
\newcommand{\e}{\mathrm e}
\newcommand{\ii}{\mathrm i}

\newcommand{\estate}[2]{\psi_{#1}^{(#2)}}
\newcommand{\enrg}[2]{E_{#1}^{(#2)}}
\usepackage[normalem]{ulem}
\newcommand{\green}[1]{#1}
\newcommand{\Green}[1]{{\color{ForestGreen}{#1}}}

 \newcommand{\blue}[1]{{\color{blue}{#1}}}
\newcommand{\red}[1]{#1}
\newcommand{\pur}[1]{#1}

\usepackage{braket}
\usepackage{bm}
\date{\today}
\usepackage[english]{babel}

\renewcommand{\blue}[2]{#2}
\renewcommand{\color}[2]{#2}

\usepackage{babel}

\makeatother

\usepackage{babel}
\begin{document}
\title{Quantum system dynamics with a weakly nonlinear Josephson junction
bath}
\author{Jing Yang}
\affiliation{Department of Physics and Astronomy, University of Rochester, Rochester,
NY 14627, USA}
\author{\'Etienne Jussiau}
\affiliation{Department of Physics and Astronomy, University of Rochester, Rochester,
NY 14627, USA}
\author{Cyril Elouard}
\affiliation{Department of Physics and Astronomy, University of Rochester, Rochester,
NY 14627, USA}
\author{Karyn Le Hur}
\affiliation{CPHT, CNRS, Institut Polytechnique de Paris, Route de Saclay, 91128
Palaiseau, France}
\author{Andrew N. Jordan}
\affiliation{Department of Physics and Astronomy, University of Rochester, Rochester,
NY 14627, USA}
\affiliation{Institute for Quantum Studies, Chapman University, 1 University Drive,
Orange, CA 92866, USA}
\begin{abstract}
We investigate the influence of a weakly nonlinear Josephson bath consisting
of a chain of Josephson junctions \Green{on the dynamics of a small quantum system (LC oscillator). Focusing on the regime where the charging
energy is the largest energy scale, we perturbatively calculate the
correlation function of the Josephson bath to the leading order in the Josephson energy divided by the charging energy while keeping the cosine potential exactly.}
When the variation of the charging energy along the chain ensures
fast decay of the bath correlation function, the dynamics of the LC
oscillator that is weakly and capacitively coupled to the Josephson
bath can be solved through the Markovian master equation. We establish
a duality relation for the Josephson bath between the regimes of large
charging and Josephson energies respectively. The results can be applied
to cases where the charging energy either is nonuniformly engineered
or disordered in the chain. Furthermore, we find that the Josephson
bath may become non-Markovian when the temperature is increased beyond
the zero-temperature limit in that the bath correlation function gets
shifted by a constant and does not decay with time.
\end{abstract}
\maketitle

\section{Introduction}

Realistic quantum systems are inevitably interacting with their surrounding environment, which induces decoherence and dissipation. In these situations, one is usually only interested in the dynamics of the primary system. Therefore some procedure that traces out the environmental degrees of freedom is required. In the past years, numerous approaches have been developed to reach this goal, including the Markovian equation developed by Gorini--Kossakowski--Sudarshan and Lindblad independently,\citep{gorini1976completely,lindblad1976onthe,CohenAtomPhoton,breuer2007thetheory} stochastic Schrödinger equations,\citep{diosi1998nonmarkovian,stockburger1999stochastic,stockburger2002exact,orth2013nonperturbative,lehur2018drivendissipative} the quantum Langevin equation,\citep{gardiner2004quantum} the Feynman--Vernon influence functional techniques,\citep{feynman1963thetheory,caldeira1983pathintegral,leggett1987dynamics,leggett1995erratum,weiss2012quantum} nonequilibrium Green's functions initiated by Schwinger,\citep{schwinger1961brownian} further developed by Keldysh\citep{keldysh1965diagram} and Kadanoff--Baym.\citep{kadanoff1994quantum} For a recent overview of the various approaches, see the review article by de Vega and Alonso.\citep{devega2017dynamics}

These sophisticated techniques, on which most of previous works have focused typically, assume either a harmonic bath (see the discussion in Breuer and Petruccione's well-known textbook\citep{breuer2007thetheory}) or an anharmonic bath whose baseline is harmonic.\citep{makri1999thelinear,hu1993quantum} Situations where a harmonic bath is coupled to a nonlinear quantum system have been widely studied within this framework. It has been shown that coupling to a nonlinear degree of freedom in the small quantum system enabled to probe many-body effects. The advent of circuit QED\citep{blais2020circuit} has given rise to recent experimental efforts in this direction.\citep{panyukov1988quantum,niemczyk2010circuit,bourassa2012josephsonjunctionembedded,nigg2012blackbox,lehur2012kondoresonance,lee2012noncanonical,moix2012equilibriumreduced,peropadre2013nonequilibrium,goldstein2013inelastic,weissl2015kerrcoefficients,forn-diaz2017ultrastrong,yoshihara2017superconducting,bosman2017multimode,magazzu2018probing,leger2019observation,martinez2019atunable,forn-diaz2019ultrastrong} 
Typically, the nonlinear small quantum system can be either represented by a two-level quantum system, \citep{leggett1987dynamics,leggett1995erratum,panyukov1988quantum,niemczyk2010circuit,lee2012noncanonical,moix2012equilibriumreduced,lehur2012kondoresonance,peropadre2013nonequilibrium,forn-diaz2017ultrastrong,yoshihara2017superconducting,bosman2017multimode,magazzu2018probing,forn-diaz2019ultrastrong} which characterizes the low-energy effective physics of a particle tunneling into a double-well potential or a Josephson junction, \citep{panyukov1988quantum,bourassa2012josephsonjunctionembedded,nigg2012blackbox,leger2019observation,martinez2019atunable} which is a naturally present source of nonlinearity in circuit-QED based qubits. This type of system has been studied extensively, ranging from the weak coupling regime,\citep{panyukov1988quantum} where the junction parameters obtain a small renormalization, to the strong coupling regime,\citep{niemczyk2010circuit,nigg2012blackbox,lehur2012kondoresonance,peropadre2013nonequilibrium,forn-diaz2017ultrastrong,yoshihara2017superconducting,bosman2017multimode,magazzu2018probing,yoshihara2018inversion,leger2019observation,martinez2019atunable,forn-diaz2019ultrastrong,jussiau2019signature,jussiau2020multiple} where an appreciable Lamb shift is produced.

Conversely, the influence of an anharmonic environment on a small quantum system has seldom been investigated. 
However, the experimental study of open quantum systems with exotic properties, beyond the usual assumption of a harmonic bath, is nowadays possible owing to advances in quantum bath engineering.\citep{murch2012cavity} There are previously works focuing on spin bath, summarized in the review article.\citep{prokof2000theory} Here, we investigate another type of nonharmonic bath, which consisiting of an array of Josephson junctions. Motivated by further technological progress,\citep{kuzmin2019quantum} we analyze in this article the dynamics of quantum system coupled to a weakly nonlinear one-dimensional Josephson junction array (JJA). The regime where the Josephson energy is the largest energy scale in play can be treated within the harmonic approximation, and the JJA then behaves as a set of harmonic oscillators to leading order. The next-to-leading-order nonlinear behavior can be well-characterized by the $\lambda\varphi^{4}$ type of nonlinearity in the usual language of anharmonic oscillators,\citep{bourassa2012josephsonjunctionembedded,weissl2015kerrcoefficients,hsiang2020nonequilibrium,yang2020nonequilibrium} and the Kerr effect associated to such a nonlinearity has already been analyzed.\citep{weissl2015kerrcoefficients} (Here $\varphi$ denotes the Josephson junction phase difference, one of the two quadratures of the oscillators with the charge difference.) In this work, we will be focusing on the opposite limit where the charging energy is much larger the Josephson energy and the temperature so that the full nonlinearity of the cosine Josephson potential must be taken into account.

Although the strong-coupling regime brings along interesting physics, we will consider here that the JJA is weakly coupled to the quantum system as a first step to probe the nonlinear environment. We compute the JJA bath correlation function to leading order in the regime of large charging energy using time-independent degenerate perturbation theory. We show that, when the nonlinear term is present and the distribution of the Josephson junction parameters is properly engineered or presents sufficient disorder, the JJA correlation function decays rapidly so that the chain behaves as a Markovian bath. The Markovianity of the JJA bath provides a route to find the dynamics of the primary quantum system within the framework of the Gorini--Kossakowski--Sudarshan--Lindblad (GKSL) master equation. To leading order, we find that the JJA bath correlation function mimics that of a harmonic bath at zero temperature, with an effective spectral density which depends on the distribution of the various junction parameters. Since the harmonic bath can be emulated by the leading order approximation of the JJA bath in the large Josephson energy regime, we establish a duality relation for the JJA in the large charging energy and the large Josephson energy regimes, which yields exactly the same coarse-grained dynamics for the small system. However, when temperature is increased beyond the zero temperature limit, we show that the dynamics of the small system becomes non-Markovian as the JJA correlation function acquires a time-independent shift that we connect to the physics of the free rotors, modeling the leading-order behaviour of the JJA.

This paper is organized as follows: In Sec.~\ref{sec:Model}, we derive the Hamiltonian of the JJA weakly coupled to an LC circuit from the standard procedure of circuit quantization. In Sec.~\ref{sec:Correlation-function}, we compute the correlation function of the JJA in the large charging energy regime from time-independent perturbation theory. This result can also be derived using the Matsubara formalism as detailed in App.~\ref{subsec:matusbara}. In Sec.~\ref{sec:non-Markovian}, we discuss the temperature driven non-Markovian effects. In Sec.~\ref{sec:DecayLS}, we compute the Lamb shift and decay rate of the damped LC oscillator within the GKSL master equation framework using the correlation function obtained in Sec.~\ref{sec:Correlation-function}. Furthermore, we also discuss the bath duality relation between the large charging energy and large Josephson energy regimes in the zero-temperature limit. We give three specific examples in Sec.~\ref{sec:Examples}. We summarize our findings and discuss future directions in Sec.~\ref{sec:Discussions}.

\section{\label{sec:Model}The model and its physical implementation}

\subsection{General description of the model}

\begin{figure*}
\begin{picture}(450,200) 
\begin{centering}
\put(50,100){\includegraphics[scale=0.45]{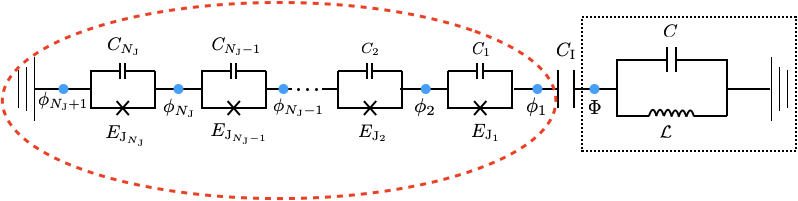}} 
\par\end{centering}
\begin{centering}
\put(50,0){\includegraphics[scale=0.42]{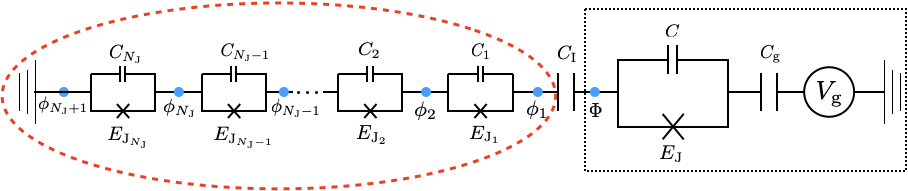}} 
\par\end{centering}
\put(25,180){\footnotesize{}{}{}{}(a)} \put(25,80){\footnotesize{}{}{}{}(b)}
\end{picture}

\caption{\label{fig:setup} (a) An LC oscillator in the right-hand side and
(b) a single Cooper-pair box in the right-hand side weakly coupled
to a one-dimensional JJA in the left-hand side of both figures. The
charging and Josephson energies are assumed to vary across the chain
in such a way that the JJA bath correlation function decays rapidly.
\red{Here, the stray capacitances between the islands and the ground
are neglected, different from the extensively studied geometry of
JJA in the literature\citep{glazman1997new,houzet2019microwave},
where the capacitances between neighboring islands are neglected but
the stray capacitances are kept nonzero.} }
\end{figure*}

Throughout this paper, we shall set $\hbar=k_{\mathrm{B}}=1$. The
Hamiltonian of the JJA, which we will derive from first principles
shortly, is 
\begin{equation}
H_{\mathrm{B}}=\sum_{\alpha}E_{\CC\alpha}N_{\alpha}^{2}-\sum_{\alpha}E_{\mathrm{J}\alpha}\cos\varphi_{\alpha},\label{eq:HB-general}
\end{equation}
where $E_{\CC\alpha}\equiv2e^{2}/C_{\alpha}$ and $E_{\mathrm{J}\alpha}$
respectively are the charging and Josephson energy for junction~$\alpha$,
while $\varphi_{\alpha}$ and $N_{\alpha}$ are canonically conjugated
variables, i.e., 
\begin{equation}
[\varphi_{\alpha},N_{\beta}]=\ii\delta_{\alpha\beta}.\label{eq:phi-Q-comm}
\end{equation}
The JJA is capacitively coupled to the quantum system of interest.
The system Hamiltonian is denoted by $H_{\mathrm{S}}$, while the
coupling Hamiltonian is written as 
\begin{equation}
H_{\mathrm{I}}=S\sum_{\alpha}g_{\alpha}N_{\alpha}.\label{eq:HI-general}
\end{equation}
In principle the system Hamiltonian $H_{\mathrm{S}}$ and the system
coupling operator $S$ can be chosen arbitrarily. We will now show
that when the system is an LC circuit capacitively coupled to the
JJA as shown in Fig.~\ref{fig:setup}(a), then the Hamiltonian of
the system and interaction terms are 
\begin{align}
 & H_{\mathrm{S}}=\frac{\mathcal{Q}^{2}}{2\mathcal{C}}+\frac{\Phi^{2}}{2\mathcal{L}},\label{eq:HS}\\
 & H_{\mathrm{I}}=\mathcal{Q}\sum_{\alpha=1}^{N_{\mathrm{J}}}g_{\alpha}N_{\alpha}.\label{eq:HI}
\end{align}
\green{Here $\mathcal{C}$ is the capacitance of the LC oscillator
which is renormalized due to the coupling to the JJA, $\mathcal{L}$
is the inductance of the LC oscillator and $N_{\mathrm{J}}$ is the
number of junctions in the JJA, while the operators $\mathcal{Q}$
and $\Phi$ correspond to the charge of the capacitor and the magnetic
flux of the inductor respectively. These operators are canonically
conjugated,} 
\begin{equation}
[\Phi,\mathcal{Q}]=\ii.\label{eq:Phi-calQ-comm}
\end{equation}

\subsection{Derivation of the Hamiltonian}

With the quantization procedure for mesoscopic circuits reviewed by
Devoret,\citep{devoret1995quantum} one can write the Lagrangian~$L_{\mathrm{tot}}$
of the circuit in Fig.~\ref{fig:setup}(a) as follows: 
\begin{align}L_{\mathrm{tot}}= & \frac{1}{2}\sum_{\alpha=1}^{N_{\JJ}}C_{\alpha}\Phi_{0}^{2}(\dot{\theta}_{\alpha}-\dot{\theta}_{\alpha+1})^{2}+\sum_{\alpha=1}^{N_{\JJ}}E_{\JJ{\alpha}}\cos\left(\theta_{\alpha}-\theta_{\alpha+1}\right) \nonumber \\
 & +\frac{1}{2}C_{\mathrm{I}}(\Phi_{0}\dot{\theta}_{1}-\dot{\Phi})^{2}+\frac{1}{2}C\dot{\Phi}^{2}-\frac{1}{2\mathcal{L}}\Phi^{2},\label{eq:L-tot}
\end{align}
where \red{$\theta_{\alpha}$ is the phase at the superconducting
island $\alpha$} and we set the ground at the extremity of the chain
such that $\theta_{N_{\mathrm{J}}+1}=0$ and $\Phi_{0}=1/(2e)$ is the
magnetic flux quantum. The charge operator ${\tilde{N}}_{\alpha}$
canonically conjugated to the phase operator $\theta_{\alpha}$ is for
$\alpha=1$ 
\begin{equation}
2e{\tilde{N}}_{1}\equiv\frac{1}{\Phi_{0}}\frac{\partial L_{\mathrm{tot}}}{\partial\dot{\theta}_{1}}=C_{1}\Phi_{0}(\dot{\theta}_{1}-\dot{\theta}_{2})+C_{\mathrm{I}}(\Phi_{0}\dot{\theta}_{1}-\dot{\Phi}),\label{eq:calN1-def}
\end{equation}
and for $\alpha>1$ 
\begin{equation}
2e{\tilde{N}}_{\alpha}\equiv\frac{1}{\Phi_{0}}\frac{\partial L_{\mathrm{tot}}}{\partial\dot{\theta}_{\alpha}}=C_{\alpha-1}\Phi_{0}(\dot{\theta}_{\alpha}-\dot{\theta}_{\alpha-1})+C_{\alpha}\Phi_{0}(\dot{\theta}_{\alpha}-\dot{\theta}_{\alpha+1}).\label{eq:calN-alpha-def}
\end{equation}
Upon making the following change of variables 
\begin{equation}
\varphi_{\alpha}=\theta_{\alpha}-\theta_{\alpha+1},
\end{equation}
the degrees of freedom for the JJA in the total Lagrangian in Eq.~(\ref{eq:L-tot})
become noninteracting, i.e., 
\begin{equation}
\begin{aligned}[t]L_{\mathrm{tot}}= & \frac{1}{2}\sum_{\alpha=1}^{N_{\JJ}}C_{\alpha}\Phi_{0}^{2}\dot{\varphi}_{\alpha}^{2}+\sum_{\alpha=1}^{N_{\JJ}}E_{\JJ{\alpha}}\cos\varphi_{\alpha}\\
 & +\frac{1}{2}C_{\mathrm{I}}(\Phi_{0}\sum_{\alpha=1}^{N_{\JJ}}\dot{\varphi}_{\alpha}-\dot{\Phi})^{2}+\frac{1}{2}C\dot{\Phi}^{2}-\frac{1}{2\mathcal{L}}\Phi^{2},\label{eq:L-tot-noninteracting}
\end{aligned}
\end{equation}
where we have used the fact that $\theta_{1}=\sum_{\alpha=1}^{N_{\mathrm{J}}}\varphi_{\alpha}$.
The charge operator canonically conjugated to the phase operator $\varphi_{\alpha}$
is 
\begin{equation}
2eN_{\alpha}\equiv\frac{1}{\Phi_{0}}\frac{\partial L_{\mathrm{tot}}}{\partial\dot{\varphi}_{\alpha}}=C_{\alpha}\Phi_{0}\dot{\varphi}_{\alpha}+C_{\mathrm{I}}(\Phi_{0}\sum_{\alpha=1}^{N_{\mathrm{J}}}\dot{\varphi}_{\alpha}-\dot{\Phi}).\label{eq:N-def}
\end{equation}
According to Eqs.~(\ref{eq:calN1-def}),~(\ref{eq:calN-alpha-def}),
and~(\ref{eq:N-def}), the charge operators ${\tilde{N}}_{\alpha}$
and $N_{\alpha}$ are related to each other through the following
relation: 
\begin{equation}
N_{\alpha}=\sum_{\beta=1}^{\alpha}{\tilde{N}}_{\beta}.\label{eq:N-calN}
\end{equation}
Clearly from this equation, $N_{\alpha}$ is a nonlocal charge operator,
i.e., the sum of all the charge operators from the preceding islands.
It is for this reason we shall call $N_{\alpha}$ and $\varphi_{\alpha}$
nonlocal variables and ${\tilde{N}}_{\alpha}$ and $\theta_{\alpha}$
local variables. To gain some intuition about Eq.~(\ref{eq:N-calN}),
we can relate the eigenstates of $N_{\alpha}$ to ${\tilde{N}}_{\alpha}$.
It is clear that the eigenstate of the nonlocal charge operators~$\{N_{\alpha}\}$,
$\ket{n_{1},\,n_{2},\,\cdots,\,n_{N_{\JJ}}}$ corresponds to the local
charge state $\ket{n_{1},\,n_{2}-n_{1},\,\cdots,\,n_{N_{\JJ}}-n_{N_{\JJ}-1}}_{\mathrm{loc}}$,
which is an eigenstate of the local charge operators~$\{{\tilde{N}}_{\alpha}\}$.
\red{Therefore, we observe that a nonlocal charge excitation corresponds
to a pair of local charge excitations with charge quantum number $+1$
and $-1$ respectively.} In what follows, we shall work with the
nonlocal operators instead of the local ones since it will simplify
the calculation dramatically.

Hereafter, we will focus on the weak-coupling regime, corresponding
to situations situations where $C_{\mathrm{I}}$ is small. In such
cases, it is natural to assume that the bath degrees of freedom are
only negligibly affected by the coupling to the LC oscillator. This
assumption is in the same spirit as the celebrated Born approximation,
widely used in the context of open quantum systems, which we apply
to this setup in Sec.~\ref{sec:DecayLS}. Therefore the \red{second
term in the last equation of Eq.~(\ref{eq:N-def}), due to the coupling
to the LC oscillator,} is ignored, which yields 
\begin{equation}
2eN_{\alpha}\approx C_{\alpha}\Phi_{0}\dot{\varphi}_{\alpha}.\label{eq:Q-approx}
\end{equation}

As regards the dynamics of the LC oscillator, we find 
\begin{equation}
\mathcal{Q}\equiv\frac{\partial L_{\mathrm{tot}}}{\partial\dot{\Phi}}=(C+C_{\mathrm{I}})\dot{\Phi}-C_{\mathrm{I}}\Phi_{0}\sum_{\alpha=1}^{N_{\mathrm{J}}}\dot{\varphi}_{\alpha}.\label{eq:calQ-def}
\end{equation}
Then, we straightforwardly obtain 
\begin{equation}
H_{\mathrm{tot}}=2e\sum_{\alpha=1}^{N_{\mathrm{J}}}N_{\alpha}\Phi_{0}\dot{\varphi}_{\alpha}+\mathcal{Q}\dot{\Phi}-L_{\mathrm{tot}}=H_{\mathrm{B}}+H_{\mathrm{I}}+H_{\mathrm{S}},\label{eq:H}
\end{equation}
where $H_{\mathrm{B}}$, $H_{\mathrm{I}}$, and $H_{\mathrm{S}}$ are
defined in Eqs.~(\ref{eq:HB-general}),~(\ref{eq:HI-general}), and~(\ref{eq:HS})
respectively, and 
\begin{align}
 & \mathcal{C}=C+C_{\mathrm{I}},\label{eq:calC}\\
 & g_{\alpha}=\frac{2eC_{\mathrm{I}}}{CC_{\alpha}}=\frac{\varepsilon_{\mathrm{I}}E_{\CC\alpha}}{e},\label{eq:g-alpha}
\end{align}
\green{where $\varepsilon_{\mathrm{I}}\equiv C_{\mathrm{I}}/C$ characterizes
the coupling strength of the LC oscillator to the JJA.}

According to Eq.~(\ref{eq:Q-approx}), $Q_{\alpha}\Phi_{0}=N_{\alpha}$
and $\varphi_{\alpha}$ are canonically conjugated variables and therefore
will satisfy Eq.~(\ref{eq:phi-Q-comm}) after being promoted to operators.
Similarly, Eq.~(\ref{eq:Phi-calQ-comm}) follows from Eq.~(\ref{eq:calQ-def}).

When the system is a qubit implemented by a Cooper pair box, as shown
in Fig.~\ref{fig:setup}(b), one can show going through a similar
procedure that $H_{\mathrm{S}}=\epsilon\sigma_{x}/2+\omega_{0}\sigma_{z}/2$
and $H_{\mathrm{I}}=\sigma_{z}\sum_{\alpha}g_{\alpha}N_{\alpha}$.\red{
In the Cooper pair box, $\sigma_{z}$ represents the two-charge states
close to a degeneracy point, $\omega_{0}$ is controlled by $V_{\mathrm{g}}$
and measures deviations from the resonance for the two charge states,
and $\epsilon$ is related to the Josephson energy $\EJ$ in Fig.~\ref{fig:setup}(b).}
In what follows, we shall focus on the case where the system is an
LC oscillator, as shown in Fig.~\ref{fig:setup}(a)---except when
it comes to the derivation of the JJA bath correlation function where
the details of the primary system do not come into play.


In the LC circuit case, using the canonical commutation relation in
Eq.~(\ref{eq:Phi-calQ-comm}), one can define the creation and annihilation
operators 
\begin{align}
 & b^{\dagger}=\sqrt{\frac{\mathcal{C}\omega_{0}}{2}}\left(\Phi-\frac{\ii\mathcal{Q}}{\mathcal{C}\omega_{0}}\right),\\
 & b=\sqrt{\frac{\mathcal{C}\omega_{0}}{2}}\left(\Phi+\frac{\ii\mathcal{Q}}{\mathcal{C}\omega_{0}}\right),
\end{align}
where 
\begin{equation}
\omega_{0}\equiv1/\sqrt{\mathcal{L}\mathcal{C}}\label{eq:omega0-def}
\end{equation}
refers to the plasma frequency in the LC system. Hence, Eqs.~(\ref{eq:HS})
and~(\ref{eq:HI}) become 
\begin{align}
 & H_{\mathrm{S}}=\omega_{0}\left(b^{\dagger}b+\frac{1}{2}\right),\\
 & H_{\mathrm{I}}=-\ii\sqrt{\frac{\mathcal{C}\omega_{0}}{2}}(b-b^{\dagger})\sum_{\alpha}g_{\alpha}N_{\alpha}.
\end{align}

\section{\label{sec:Correlation-function}Correlation function of the JJA
bath}

\subsection{The central role of the bath correlation function}

It is widely known that the dynamics of a ``small'' quantum system
(an LC oscillator here) coupled to a large bath (the Josephson junction
chain here) can be resolved thanks to the celebrated GKSL master equation,\citep{gorini1976completely,lindblad1976onthe,CohenAtomPhoton,breuer2007thetheory}
which has been proven to yield the most general Markovian evolution
preserving the fundamental properties of the system density matrix.
This formalism can be applied to a wide variety of situations, provided
specific conditions are met by the system-bath coupling. Namely, to
ensure that the master equation is Markovian, it is usually hypothesized
that the coupling to the primary system negligibly influences the
bath dynamics, which is typically the case when they are weakly coupled.
In this context, the total system-bath density matrix is assumed to
be factorized at all times, $\rho(t)=\rho_{\mathrm{S}}(t)\otimes\rho_{\mathrm{B}}$,
where the (constant) bath density matrix~$\rho_{\mathrm{B}}$ corresponds
to the canonical distribution with inverse temperature~$\beta$,
\begin{equation}
\rho_{\mathrm{B}}=\frac{\e^{-\beta H_{\mathrm{B}}}}{\Tr\e^{-\beta H_{\mathrm{B}}}}.\label{ThState}
\end{equation}
\Green{While the interaction between the system and the bath induces correlations between them, the evolution of such ansatz is a good approximation of the system dynamics coarse-grained over a timescale much larger than the correlation time of the bath, i.e., the timescale over which the correlations between the system and the bath decay.\citep{CohenAtomPhoton} This coarse-graining time must also be smaller than the typical decay time for the system, which is only possible if the correlation time is much smaller than this decay time. This coarse-grained dynamics is captured by the GKSL master equation.
The existence of a finite, and small, correlation time for the bath can be verified by considering the appropriate correlation function of the bath, a central quantity for the study of Markovian open quantum systems}. The choice
of the relevant bath correlation function to analyze is dictated by the
reservoir observable involved in the coupling Hamiltonian, which is
the nonlocal charge operator $N_{\alpha}$ here. A brief derivation
of the GKSL master equation highlighting the importance of the reservoir
correlation function is given in App.~\ref{sec:GKSL} and \Green{the GKSL master equation is given by Eq.~(\ref{eq:GKSL}) with the coefficients given by Eqs.~(\ref{eq:coeff-delta-omg}), ~(\ref{eq:coeff-kappa-omg}) and~(\ref{eq:coeff-Gamma-omg}) }.

Furthermore, the GKSL formalism is designed for situations where
\Green{the coupling Hamiltonian satisfies}
\footnote{If it is not the case, then one can always redefine the system and coupling
Hamiltonians~$H_{\mathrm{S}}$ and~$H_{\mathrm{I}}$ such that Eq.~(\ref{eq:normalization})
holds, see Refs.~{[}\onlinecite{CohenAtomPhoton}{]} and~{[}\onlinecite{breuer2007thetheory}{]}
for more details.} 
\begin{equation}
\Tr_{\mathrm{B}}[\rho_{\mathrm{B}}H_{\mathrm{I}}]=0,\label{eq:normalization}
\end{equation}
\Green{where the trace is taken over the bath degrees of freedom only. As such, since the coupling Hamiltonian~$H_\mathrm I$ acts on both the system and bath, the right-hand side of Eq.~(\ref{eq:normalization}) has to be understood as an operator acting on the system Hilbert space.}
When studying an LC oscillator coupled to a JJA, the coupling Hamiltonian
is given in Eq.~(\ref{eq:HI})\Green{, which yields}
\begin{equation}
\Tr_{\mathrm{B}}[\rho_{\mathrm{B}}H_{\mathrm{I}}]=\mathcal{Q}\sum_{\alpha}g_{\alpha}\braket{N_{\alpha}}.
\end{equation}
Here $\langle\bullet\rangle$ denotes the average over the thermal
state~$\rho_{\mathrm{B}}$ in Eq.~(\ref{ThState}).

It turns out that the expectation value~$\braket{N_{\alpha}}$ vanishes
as a consequence of charge conjugation symmetry. This property concerns
single junctions so we can temporarily drop the subscript~$\alpha$.
Let us define the charge conjugation operator~$\mathscr{C}$ such
that $\mathscr{C}\ket n=\ket{-n}$. $\mathscr{C}$ is unitary and
Hermitian: $\mathscr{C}^{\dagger}=\mathscr{C}^{-1}=\mathscr{C}$.
It is straightforward to check that the number operator is odd under
charge conjugation: $\mathscr{C}N\mathscr{C}=-N$. \Green{However, because of the commutation relation $[\varphi,\,N]=\ii$, one can easily show that  $\e^{\pm \ii \varphi} \ket{n}=\ket{n\mp 1}$ and therefore $\mathscr{C}(\cos \varphi) \mathscr{C}=\cos \varphi$.} So we conclude that the junction
Hamiltonian $H=\EC N^{2}-\EJ\cos\varphi$ is invariant under this
transformation: $\mathscr{C}H\mathscr{C}=H$. We then have 
\begin{equation}
\braket{N}=\frac{1}{Z}\Tr(\e^{-\beta H}N)=\frac{1}{Z}\Tr(\mathscr{C}\e^{-\beta H}\mathscr{C}\mathscr{C}N\mathscr{C})=-\braket{N}.
\end{equation}
It is then clear that $\braket{N}=0$, which proves that Eq.~(\ref{eq:normalization})
is satisfied here. The GKSL formalism can then be applied and the
relevant correlation function for the JJA bath is given by 
\begin{equation}
\Gamma(t)=\sum_{\alpha}g_{\alpha}^{2}G_{\alpha}(t),\label{eq:Gamma-t}
\end{equation}
where $G_{\alpha}(t)$ is the correlation function for a single Josephson
junction, 
\begin{equation}
G_{\alpha}(t)=\braket{N_{\alpha}(t)N_{\alpha}(0)}.
\end{equation}
Hereafter time-dependent operators indicate the interaction picture
(Heisenberg picture with respect to Hamiltonian~$H_{\mathrm{B}}$).

\subsection{The large Josephson energy limit: The harmonic approximation}

In what follows, we shall detail the calculation of the single-junction
correlation function $G_{\alpha}(t)$ and then take the continuum
limit to obtain the bath correlation function $\Gamma(t)$. The main
aim of this paper is to discuss the large charging energy regime.
However, before we head toward this end, let us first detour to discuss
the extensively studied large Josephson energy regime, that is $E_{\CC\alpha}\ll E_{\mathrm{J}\alpha}$
and $\beta^{-1}\ll E_{\mathrm{J}\alpha}$, where the harmonic bath
approximation can be applied.\citep{weissl2015kerrcoefficients,bourassa2012josephsonjunctionembedded}
In this regime, thermal excitations are near a fixed minimum of the
cosine potential and the effect of quantum phase slips may be neglected
to leading order. In this case, the JJA bath can be approximated by
a harmonic bath such that the bath Hamiltonian in Eq.~(\ref{eq:HB-general})
becomes 
\begin{equation}
H_{\mathrm{B}}=\sum_{\alpha}\left[\frac{1}{2L_{\alpha}}\Phi_{\alpha}^{2}+\frac{1}{2}L_{\alpha}\omega_{\alpha}^{2}Q_{\alpha}^{2}\right],
\end{equation}
where $L_{\alpha}=\Phi_{0}^{2}/E_{\mathrm{J}\alpha}$ is the effective
Josephson inductance---not to be confused with the total Lagrangian~$L_{\mathrm{tot}}$---and
$\omega_{\alpha}$ is the characteristic frequency of the oscillator
given by 
\begin{equation}
\omega_{\alpha}=\frac{1}{\sqrt{L_{\alpha}C_{\alpha}}}=\sqrt{2E_{\mathrm{J}\alpha}E_{\CC\alpha}}.
\end{equation}
Even though $E_{\CC\alpha}\ll E_{\mathrm{J}\alpha}$, the characteristic
frequency~$\omega_{\alpha}$ can still take values in a wide spectrum.
Therefore, by properly controlling the variations of the charging
and Josephson energies along the JJA, one can still emulate a harmonic
bath, which is usually implemented in transmission lines. The correlation
function for a harmonic oscillator is well-known,\citep{breuer2007thetheory}
namely 
\begin{equation}
G_{\alpha}(t)=\frac{\omega_{\alpha}}{4E_{\CC\alpha}}\left[\coth\left(\frac{\beta\omega_{\alpha}}{2}\right)\cos(\omega_{\alpha}t)-\ii\sin(\omega_{\alpha}t)\right].\label{eq:Gi-harmonic}
\end{equation}
Since a genuine thermal bath has infinitely many degrees of freedoms,
we have to find the correlation function in the continuum limit. To
this end, we assume that Josephson junctions are spatially distributed
along the JJA according to the density $\nu(x)$. The JJA correlation
function is then given by 
\begin{equation}
\Gamma(t)=\int\dd{x}\nu(x)g^{2}(x)G(x,t),
\end{equation}
where the junction number index~$\alpha$ has been replaced by the
position variable~$x$, and the explicit expression for the coupling
parameter~$g(x)$ is analogous to that given in Eq.~(\ref{eq:g-alpha}).
Through a simple change of variables, it is straightforward to obtain
\begin{equation}
\Gamma(t)=\qty(\frac{\varepsilon_{\mathrm{I}}}{2e})^{2}\int\dd{\omega}J(\omega)\qty[\coth\left(\frac{\beta\omega}{2}\right)\cos(\omega t)-\ii\sin(\omega t)],\label{eq:Gamma-HC}
\end{equation}
where $J(\omega)$ is the spectral density defined as 
\begin{equation}
J(\omega)=\sum_{k}\omega\nu_{k}(\omega)\EC^{(k)}(\omega)\abs{\dv{x_{k}}{\omega}\strut(\omega)}.\label{eq:J-harmonic}
\end{equation}
Here the index~$k$ labels the different intervals~$I_{k}$ on which
$\omega(x)$ is a monotonic function. On each of these intervals,
$\omega(x)$ can be inverted and $x_{k}(\omega)$ denotes the corresponding
inverse function. From this, we define 
\begin{align}
 & \nu_{k}(\omega)\equiv\nu(x_{k}(\omega)),\\
 & \EC^{(k)}(\omega)\equiv\EC(x_{k}(\omega)).
\end{align}
Hereafter, we will focus on the low-temperature regime where $\beta^{-1}\ll\omega(x)$
for (almost) all $x$. In that case, Eq.~(\ref{eq:Gamma-HC}) becomes
\begin{equation}
\Gamma(t)=\left(\frac{\varepsilon_{\mathrm{I}}}{2e}\right)^{2}\int_{0}^{\infty}\dd{\omega}J(\omega)e^{-\mathrm{i}\omega t}.\label{eq:Gamma-EJ-0T}
\end{equation}
It should be noted that in the large $\EJ$ regime where $\EC(x)\ll\EJ(x)$,
the condition $\beta^{-1}\ll\omega(x)$ implies that $\beta^{-1}\ll\EJ(x)$.

\red{For the extensively studied geometry of JJA, where the stray
capacitances are nonvanishing but the capacitances between neighboring
islands are neglected,\citep{glazman1997new,houzet2019microwave}
in large Josephson energy and zero-temperature limits, one can show
that $\Gamma(t)$ decays as power-law according to Eq.~(\ref{eq:Gamma-EJ-0T}),
due to the linear dispersion relation of sound modes characterizing
the lead-order effects of the JJA. However, in our setup shown in
Fig.~\ref{fig:setup}, the behavior of $\Gamma(t)$ is controlled
by the distribution of the junction parameters in real space, rather
than the sound modes in the momentum space. Therefore, in our setup
$\Gamma(t)$ does not necessarily decay as power law in large Josephson
energy and zero-temperature limits.}

\subsection{The large charging energy limit}

In the remainder of this article, we focus on the large charging energy
limit where 
\begin{align}
 & E_{\mathrm{J}\alpha}\ll E_{\CC\alpha},\,\sqrt{E_{\CC\alpha}\delta\EC},\label{eq:EJ-small}\\
 & \beta^{-1}\ll\Delta_{\alpha}\equiv\frac{E_{\CC\alpha}}{-\ln\lambda_{\alpha}},\label{eq:LowT}
\end{align}
where $\lambda_{\alpha}\equiv E_{\mathrm{J}\alpha}/E_{\CC\alpha}\ll1$
and $\delta\EC$ is the width of the effective spectral density of
the Josephson bath {[}defined subsequently in Eq.~(\ref{eq:JEC-def}){]},
which represents the characteristic frequency of the bath. In this
regime, the physics is dominated by capacitors, which are quantum
mechanically equivalent to free rotors from Eq.~(\ref{eq:HB-general}).
\red{The ground state fixes the charge but the phase variable can
fluctuate accordingly.} Thermal excitation in a junction is negligible
so that each junction stays in the exact ground state of the total
Hamiltonian to the leading order. 
Based on this intuition, we can evaluate $G_{\alpha}(t)$ to second
order in $\lambda_{\alpha}$ at low temperature perturbatively with
either the time-independent degenerate perturbation theory\citep{CohenTannoudji2}
or the Matsubara imaginary time formalism.\citep{das1997finitetemperature,mahan2013many}
\green{However, it is unclear whether results obtained through the
Matsubara formalism hold at very low temperatures. The time-independent
perturbation method does not suffer from such limitation, which is
why we focus on this method in the main text, with further details
given in App.~\ref{App_PerturbTheory}. The derivation of the correlation
function using the Matsubara formalism can be found in App.~\ref{subsec:matusbara}.}
When Eqs.~(\ref{eq:EJ-small}) and~(\ref{eq:LowT}) hold, the single-junction
correlation function reads 
\begin{equation}
G_{\alpha}(t)=\frac{\lambda_{\alpha}^{2}}{2}\e^{-\ii E_{\CC\alpha}t}+O(\lambda_{\alpha}^{3},\e^{-\beta E_{\CC\alpha}}).\label{eq:Gi}
\end{equation}


Note that Eq.~(\ref{eq:Gi}) is dramatically different from Eq.~(\ref{eq:Gi-harmonic})
as the characteristic oscillation frequency in Eq.~(\ref{eq:Gi})
is $E_{\CC\alpha}$ instead of $\sqrt{E_{\JJ\alpha}E_{\CC\alpha}}$.
This substantial difference is due the fact that they are valid in
two opposite regimes. \green{The first frequency represents the energy
necessary to add a charge on a capacitor, while the second one defines
the plasma (resonance) frequency due to sound modes in the JJA.}

\subsubsection{Derivation of the single-junction correlation function through time-independent
perturbation method}

We now derive the correlation function in the limit of large charging
energy for a single Josephson junction so we can drop the subscript
$\alpha$. The corresponding Hamiltonian is $H=H_{\CC}+\lambda H_{\JJ}$,
where $\lambda=\EJ/\EC\ll1$ and 
\begin{align}
 & H_{\CC}=\EC N^{2},\label{eq:HC-singleJ}\\
 & H_{\JJ}=-\EC\cos\varphi.\label{eq:HJ-singleJ}
\end{align}

We consider time-independent perturbation theory treating the Josephson
Hamiltonian as a perturbation. In this framework, we compute the average
values in the correlation function explicitly using the eigenenergies
and eigenstates obtained from our perturbative calculation.

As pointed out earlier, the Hamiltonian~$H$ is invariant under charge
conjugation, $\mathscr{C}H\mathscr{C}=H$, which yields $[H,\mathscr{C}]=0$.
As such, $H$ and $\mathscr{C}$ share a common eigenbasis. Since
$\mathscr{C}^{2}=1$, the eigenvalues of $\mathscr{C}$ are $\pm1$,
which means that the corresponding eigenstates are either symmetric
or antisymmetric under charge conjugation. We consequently denote
by $\ket{\psi_{n,\pm}}$ the common eigenstates of $H$ and $\mathscr{C}$
such that 
\begin{align}
 & H\ket{\psi_{n,\pm}}=E_{n,\pm}\ket{\psi_{n,\pm}},\label{EvalEq}\\
 & \mathscr{C}\ket{\psi_{n,\pm}}=\pm\ket{\psi_{n,\pm}},
\end{align}
\green{where the index~$n$ is a positive integer that labels the
energy levels of the charging Hamiltonian~$H_{\CC}$.} Working in
this basis, the single-junction correlation function~$G(t)=\braket{N(t)N(0)}$
is expressed as 
\begin{equation}
\begin{aligned}[t]G(t)=\frac{1}{Z}\sum_{m,n,\pm}\abs{\braket{\psi_{m,\pm}|N|\psi_{n,\mp}}}^{2}\e^{-\beta E_{m,\pm}}\e^{\ii(E_{m,\pm}-E_{n,\mp})t},\end{aligned}
\label{G}
\end{equation}
with the partition function 
\begin{equation}
Z=\sum_{n,\pm}\e^{-\beta E_{n,\pm}}.
\end{equation}
One should note that, since the number operator~$N$ is odd under
charge conjugation, it only couples states of different charge parities.
This is why we only have to consider scalar products of the type~$\braket{\psi_{m,\pm}|N|\psi_{n,\mp}}$
in Eq.~(\ref{G}).

Let us now compute the eigenstates~$\ket{\psi_{n,\pm}}$ and eigenenergies~$E_{n,\pm}$
to lowest orders in $\lambda$ using time-independent perturbation
theory.\citep{CohenTannoudji2} The starting point is to write $E_{n,\pm}$
and $\ket{\psi_{n,\pm}}$ as power series in $\lambda$,
\begin{align}
 & E_{n,\pm}=\sum_{q=0}^{\infty}\lambda^{q}\enrg{n,\pm}q,\\
 & \ket{\psi_{n,\pm}}=\sum_{q=0}^{\infty}\lambda^{q}\ket{\estate{n,\pm}q}.
\end{align}
Equating the two sides of Eq.~(\ref{EvalEq})
to all orders in $\lambda$ then yields 
\begin{equation}
H_{\CC}\ket{\estate{n,\pm}q}+H_{\JJ}\ket{\estate{n,\pm}{q-1}}=\sum_{p=0}^{q}\enrg{n,\pm}{p}\ket{\estate{n,\pm}{q-p}}.\label{PerturbEvalEq}
\end{equation}
To the zeroth order in $\lambda$, we simply obtain 
\begin{equation}
H_{\CC}\ket{\estate{n,\pm}0}=\enrg{n,\pm}0\ket{\estate{n,\pm}0}.\label{EvalEq0}
\end{equation}
This means that $\enrg{n,\pm}0$ is an eigenenergy of $H_{\CC}$,
with $\ket{\estate{n,\pm}0}$ being the corresponding eigenstate.
While we can readily identify $\enrg{n,\pm}0$ with the charging energy,
$\enrg{n,\pm}0=n^{2}\EC$, the state $\ket{\estate{n,\pm}0}$ remains
undetermined at this stage. This is because all the energy levels
of the charging Hamiltonian except the ground state are two-fold degenerate,
and Eq.~(\ref{EvalEq0}) then only tells us that $\ket{\estate{n,\pm}0}$
belongs to the subspace generated by the charge states~$\ket n$
and~$\ket{-n}$. 
However, since each state~$\ket{\psi_{n,\pm}}$ has a definite charge
parity---we recall that it is an eigenstate of both $H$ and $\mathscr{C}$---,
all the corrections~$\ket{\estate{n,\pm}q}$ must have the same property.
\green{This includes the zeroth order eigenstates which must also
satisfy $\mathscr{C}\ket{\estate{n,\pm}0}=\pm\ket{\estate{n,\pm}0}$.}
For all $n>0$, we then find the appropriate choice for these states
to be 
\begin{equation}
\ket{\estate{n,\pm}0}=\ket{\chi_{n,\pm}}\equiv\frac{1}{\sqrt{2}}(\ket n\pm\ket{-n}),
\end{equation}
where $\mathscr{C}\ket{\chi_{n,\pm}}=\pm\ket{\chi_{n,\pm}}$. Hence,
each two-fold degenerate charge energy level can be further subdivided
into two eigenstates of different charge parities; we expect the degeneracy
of these states to be lifted by the Josephson Hamiltonian. Finally,
one should note that the relevant zeroth order eigenstate for $n=0$
is simply $\ket{\estate00}=\ket0$, which indicates that the ground
state $\ket{\psi_{0}}$ is even under charge conjugation. 

Using Eq.~(\ref{PerturbEvalEq}) with $q=1$, we find the corrections
to the energy to first order in $\lambda$, 
\begin{equation}
\enrg{n,\pm}1=\braket{\estate{n,\pm}0|H_{\JJ}|\estate{n,\pm}0}.
\end{equation}
The Josephson potential only couples neighboring charge states as,
for any charge states~$\ket m$ and~$\ket n$, we have 
\begin{equation}
\braket{m|H_{\JJ}|n}=-\frac{\EC}{2}(\delta_{m,n+1}+\delta_{m,n-1}).
\end{equation}
Thus, the perturbation does not yield any correction to the energies
to first order in $\lambda$, $\enrg{n,\pm}1=0$.

The first-order eigenstates are given by 
\begin{equation}
\ket{\estate{n,\pm}1}=-\sum_{m\ne n}\frac{\braket{\estate{m,\pm}0|H_{\JJ}|\estate{n,\pm}0}}{(m^{2}-n^{2})\EC}\ket{\estate{m,\pm}0}.
\end{equation}
Note that here the norm and phase of $\ket{\psi_{n,\pm}}$ have been
chosen such that $\braket{\psi_{n,\pm}|\psi_{n,\pm}}=1$ and $\braket{\estate{n,\pm}0|\psi_{n,\pm}}$
be real. To first order in $\lambda$, this imposes $\braket{\estate{n,\pm}0|\estate{n,\pm}1}=0$.
In what follows, the cases for $n=0$ or $n=1$ must be treated separately
since they involve the ground state which is clearly distinct from
the excited states as regards the structure of their leading-order
terms. Namely, we have $\ket{\estate00}=\ket0$ while, for $n>0$,
$\ket{\estate{n,\pm}0}=\ket{\chi_{n,\pm}}$. For $n=0$, we obtain
\begin{equation}
\ket{\psi_{0}}=\ket0+\frac{\lambda}{\sqrt{2}}\ket{\chi_{1,+}}+O(\lambda^{2}).
\label{psi01}
\end{equation}
For $n=1$, we find 
\begin{align}
 & \ket{\psi_{1,+}}=\ket{\chi_{1,+}}+\frac{\lambda}{6}\ket{\chi_{2,+}}+\frac{\lambda}{\sqrt{2}}\ket0+O(\lambda^{2}),\\
 & \ket{\psi_{1,-}}=\ket{\chi_{1,-}}+\frac{\lambda}{6}\ket{\chi_{2,-}}+O(\lambda^{2}).
\end{align}
Finally, for $n>1$, we have 
\begin{equation}
\ket{\psi_{n,\pm}}=\ket{\chi_{n,\pm}}+\frac{\lambda}{4n+2}\ket{\chi_{n+1,\pm}}-\frac{\lambda}{4n-2}\ket{\chi_{n-1,\pm}}+O(\lambda^{2}).\label{psin1}
\end{equation}

We now consider the second-order corrections to the energy. According
to Eq.~(\ref{PerturbEvalEq}) with $q=2$, we have 
\begin{equation}
\enrg{n,\pm}2=-\sum_{m\ne n}\frac{\abs{\braket{\estate{m,\pm}0|H_{\JJ}|\estate{n,\pm}0}}^{2}}{(m^{2}-n^{2})\EC}.
\end{equation}
For $n=0$, this yields 
\begin{equation}
E_{0}=-\frac{\lambda^{2}\EC}{2}+O(\lambda^{3}).
\end{equation}
For $n=1$, we obtain 
\begin{align}
 & E_{1,+}=\EC\qty(1+\frac{5\lambda^{2}}{12})+O(\lambda^{3}),\\
 & E_{1,-}=\EC\qty(1-\frac{\lambda^{2}}{12})+O(\lambda^{3}).
\end{align}
The Josephson Hamiltonian thus lifts the degeneracy of the first excited
eigenstates to second order in $\lambda$. Conversely, higher-energy
excited states remain degenerate to this point as, for $n>1$, we
find 
\begin{equation}
E_{n,\pm}=\EC\qty(n^{2}+\frac{\lambda^{2}}{2(4n^{2}-1)})+O(\lambda^{3}).
\end{equation}

We can now use the result obtained with the perturbation method to
compute the single-junction correlation function in Eq.~(\ref{G}).
However, the eigenenergies only appear \green{in oscillating exponentials}
multiplied by the time~$t$ or \green{in the Boltzmann factors multiplied}
the inverse temperature~$\beta$ in Eq.~(\ref{G}). This must be
accounted for \green{in subsequent calculations as one has to ensure
that energy corrections to high orders in $\lambda$ can still be
neglected. This is clearly the case if we restrict ourselves to short
times, $t\ll1/(\lambda^{2}\EC)$. This limitation to short times does
not constitute a hindrance to our analysis. Indeed, we are interested
here in situations where the correlation function for the whole chain
rapidly decays so that the Born--Markov approximations hold. The
crucial point of our calculation is to ensure that this fast decay
does happen, and we do not need to resolve the dynamics of correlations
for longer times. Therefore, we will only consider the short-time
regime~$t\ll1/(\lambda^{2}\EC)$ hereafter. Overall, our approach
provides a satisfactory description of correlations in a single junction,
so long as $\omega_{\mathrm{B}}^{-1}\ll1/(\lambda^{2}\EC)$, where
$\omega_{\mathrm{B}}^{-1}$ is the correlation time for the whole
chain. We typically estimate $\omega_{\mathrm{B}}\sim\delta\EC$,
where $\delta\EC$ is the width of the charging energy distribution
across the chain, as shown in Eq.~(\ref{eq:Gamma}) later. This results
in the constraint $\EJ\ll\sqrt{\EC\delta\EC}$ in Eq.~(\ref{eq:EJ-small}).
A more accurate estimate of $\omega_{\mathrm{B}}$ for a specific
example will be derived in Eq.~(\ref{eq:omega-B}) later. }

\green{As regards the Boltzmann factors, we will actually consider
two opposite regimes of temperature in what follows. First, we analyze
the regime of high temperatures, $\beta^{-1}\gg\lambda^{2}\EC$. This
is in the same spirit as the limitation to short times discussed above.
In this regime, terms to high orders in $\lambda$ can be neglected
in the Boltzmann factors so our perturbative results can be used.
Then, in a second time, we tackle the low-temperature regime. In this
case, we cannot estimate the Boltzmann factors with accuracy through
a series in powers of $\lambda$. However, if the temperature is low
enough, then we can only keep the contribution of the ground state to the
correlation function. This is because the Boltzmann factors corresponding
to excited states, $\e^{-\beta E_{n,\pm}}$ with $n>0$, are negligible
with respect to $\e^{-\beta E_{0}}$, which is typically the case
when $\beta^{-1}\ll\EC$. Then, the Boltzmann factors~$\e^{-\beta E_{0}}$
in the numerator and the denominator cancel each other so it is not
necessary to have the full expansion for the energy~$E_{0}$.}



\green{Let us first analyze the regime of high temperatures where
$\beta^{-1}\gg\lambda^{2}\EC$. For simplicity, we only keep terms
to leading order in $\lambda$, $E_{n,\pm}=n^{2}\EC+O(\lambda^{2})$
and $\braket{\psi_{m,\pm}|N|\psi_{n,\mp}}=m\delta_{mn}+O(\lambda)$
(recall that $N$ only couples states of different charge parities).
In this context, we find that the single-junction correlation reduces
to that of a free rotor, 
\begin{equation}
G(t)\approx\frac{2\sum_{n=1}^{\infty}n^{2}\e^{-n^{2}\beta\EC}}{1+2\sum_{n=1}^{\infty}\e^{-n^{2}\beta\EC}},\label{G_highT}
\end{equation}
where we recall that the short-time limit, $t\ll1/(\lambda^{2}\EC)$,
is considered here. The factors of 2 above illustrate that excited
states of different parities contribute equally to the correlation
function.}

\green{We now turn to} the opposite low-temperature regime where
$\beta^{-1}\ll\EC$. \Green{In this case, we only keep the terms in Eq.~(\ref{G}) whose Boltzmann factor is associated to the ground state. The correlation function then reads}
\begin{equation}
G(t)\approx\sum_{n=1}^{\infty}\abs{\braket{\psi_{0}|N|\psi_{n,-}}}^{2}\e^{-\ii(E_{n,-}-E_{0})t}.\label{eq:G-ground}
\end{equation}
Again, $N$ only couples states of different charge parities and $\ket{\psi_{0}}$
is even under charge conjugation, which is why the sum above only
features the odd eigenstates~$\ket{\psi_{n,-}}$. Moreover, since
\green{$N\ket0=0$}, we deduce that the term of zeroth order in
$\lambda$ vanishes here. To leading-order in $\lambda$, only $\ket{\psi_{1,-}}$
contributes to the sum, so that we find, for $t\ll1/(\lambda^{2}\EC)$,
\begin{equation}
G(t)\approx\frac{\lambda^{2}}{2}\e^{-\ii\EC t}.\label{G_lowT}
\end{equation}

We observe two clearly different dynamics for correlations in Eqs.~(\ref{G_highT})
and~(\ref{G_lowT}). In the high temperature regime, we find that
the correlation function is constant. This rules out the possibility
for the whole chain's correlation function to decay fast and forbids
Markovianity. Crucially, in the case of free rotors ($\lambda=0$),
the dynamics of the correlation function is always given by this constant
term so the Born--Markov approximation never applies. On the contrary,
Eq.~(\ref{G_lowT}) features an exponential oscillating in time.
We will show later that this time dependence can give rise to a rapidly
decaying correlation function for the whole chain to leading order
provided that the charging energy distribution along the chain is
appropriately chosen. This underlines the primary importance of the
Josephson potential as it \green{introduces an overlap between the
ground state~$\ket{\psi_{0}}$ and the even charge eigenstate~$\ket{\chi_{1,+}}$
as shown in} Eq.~(\ref{eq:G-ground}). This overlap gives rise to
the oscillating behavior of the correlation function, which is the
leading-order behavior in the low-temperature regime. In summary,
we observe that the Josephson junction chain can behave as a Markovian
bath at low temperature while this is no longer the case for higher
temperatures. Here, Markovianity breaks down as temperature increases
contrary to what is usually witnessed in the usual harmonic bath,
\red{where the short correlation time at high temperature guarantees
Markovianity, as one can see from Eq.~(\ref{eq:Gamma-HC}). } 

\begin{figure}
\centering
\includegraphics[width=\linewidth]{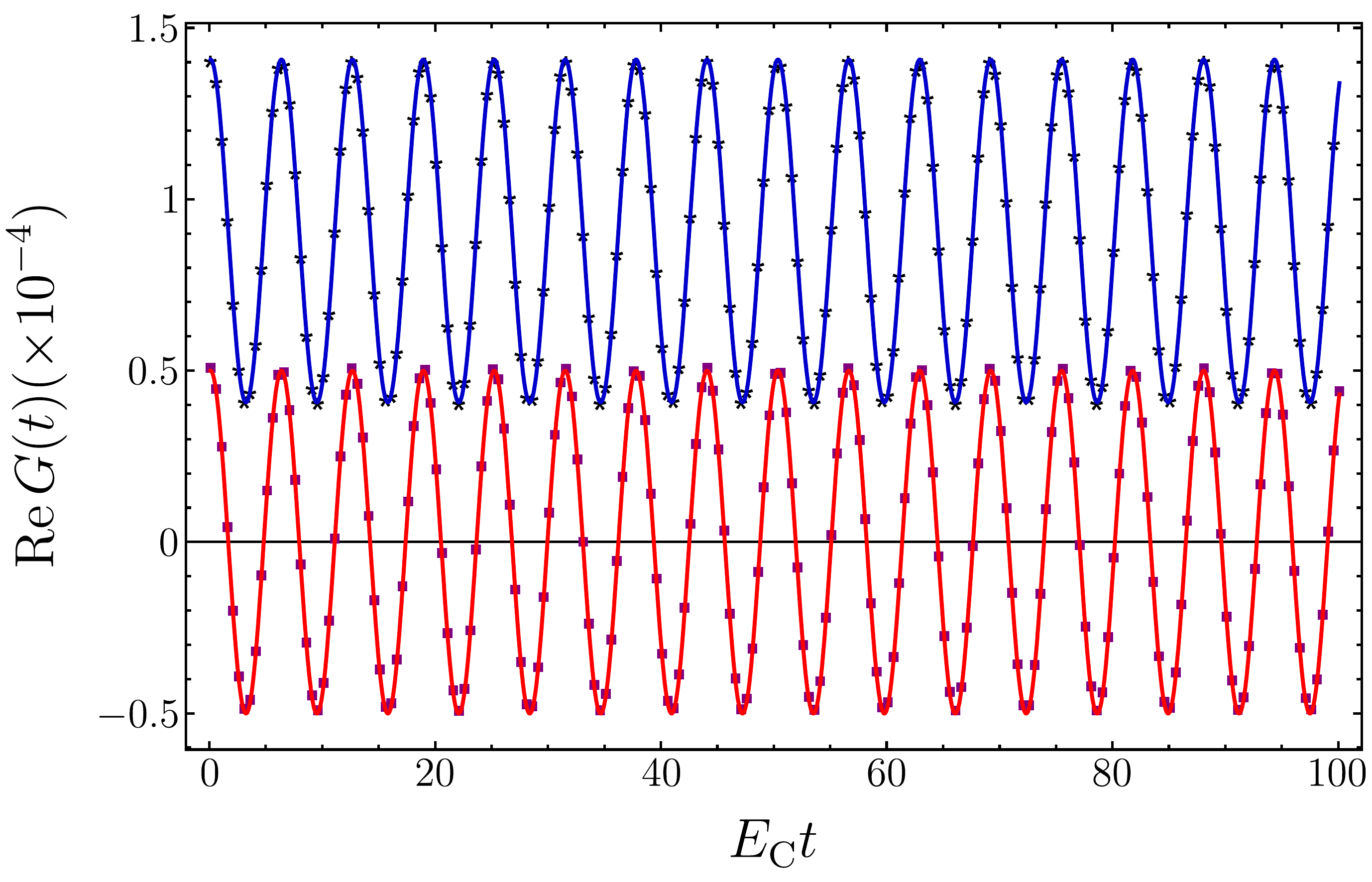} 
\caption{\label{fig:comp}Check of $G(t)$ against numerical calculation. The
value of the Josephson energy is $\EJ=0.01\EC$. The red and
blue solid lines are analytical calculations of the real parts of
$G(t)$ at temperatures $\beta^{-1}\ll\EC/\ln(\EC/\EJ)=0.2\EC$ and
$\beta^{-1}=0.1\EC$ respectively, which are plotted according to
Eq.~(\ref{G_eps}). The purple squares and black stars are the corresponding
numerical calculations of the real parts of $G(t)$ for these two
temperatures. We clearly see that the perturbative results from Eq.~(\ref{G_eps})
are in excellent agreement with the numerical results for $t\ll1/(\lambda^{2}\EC)=10^{4}/\EC$.
Furthermore, the constant offset in $G(t)$, i.e., the second term
in the right hand side of Eq.~(\ref{G_eps}), which becomes nonnegligible
at temperature $\beta^{-1}\sim\EC/\ln(\EC/\EJ)$, is also confirmed
by the numerical calculations.}
\end{figure}

In the low-temperature regime, only the ground state's contribution
to correlations have been taken into account \red{when moving from
Eq.~(\ref{G}) to Eq.~(\ref{eq:G-ground})}. When temperature is
increased, contributions from excited states must also be taken into
consideration. In doing so, the transition in the dynamics of correlation
from oscillatory to constant can be analyzed more thoroughly. For
example, when $\e^{-\beta\EC}\sim\lambda^{2}$, that is $\beta^{-1}\sim\EC/(-\ln\lambda)$,
it is necessary to include the contribution from the first excited
states into the calculation. We then obtain 
\begin{equation}
G(t)\approx\frac{\lambda^{2}}{2}\e^{-\ii\EC t}+2\e^{-\beta\EC}.\label{G_eps}
\end{equation}
This expression provides a satisfactory description of single-junction
correlation function \green{for short times, $t\ll1/(\lambda^{2}\EC)$,
and moderately low temperatures, $\lambda^{3}\EC\ll\beta^{-1}\ll\EC$.
This justifies a posteriori the form of Eq.~(\ref{eq:LowT}) refining
the regime of validity of Eq.~(\ref{G_lowT})}. Fig.~\ref{fig:comp}
shows that Eq.~(\ref{G_eps}) can characterize the dynamics for $t\ll1/(\lambda^{2}\EC)$
with excellent precision and the constant offset corresponding to
the second term in the right hand side of Eq.~(\ref{G_eps}) is confirmed by the numerics. \Green{The simulation is performed by numerically calculating $\braket{N(t)N(0)}$ taking into account the first $41$ charge eigenbasis states, namely the states~$\ket n$ for $n=-20,\dots,20$.}

More details about the perturbative calculation of the single-junction
correlation function are given in App.~\ref{App_PerturbTheory}.

\subsubsection{JJA correlation function in the continuum limit}

When temperature is further lowered such at Eq.~(\ref{eq:LowT})
is satisfied, the subleading order is much smaller than the leading
order in Eq.~(\ref{G_eps}). Therefore, it reduces to Eq.~(\ref{eq:Gi}).
In this regime, the Josephson bath correlation function obtained from
Eq.~(\ref{eq:Gi}) is 
\begin{equation}
\Gamma(t)=\sum_{\alpha}\frac{g_{\alpha}^{2}E_{\mathrm{J}\alpha}^{2}}{2E_{\CC\alpha}^{2}}\e^{-\ii E_{\CC\alpha}t}=\frac{1}{2}\left(\frac{\varepsilon_{\mathrm{I}}}{e}\right)^{2}\sum_{\alpha}E_{\mathrm{J}\alpha}^{2}\e^{-\ii E_{\CC\alpha}t},\label{eq:Gamma-discrete}
\end{equation}
where we have used the explicit expression in Eq.~(\ref{eq:g-alpha}) for
$g_{\alpha}$ in the last equality. In the continuum limit, this becomes
\begin{equation}
\Gamma(t)=\frac{1}{2}\left(\frac{\varepsilon_{\mathrm{I}}}{e}\right)^{2}\int\dd{x}\nu(x)[\EJ(x)]^{2}\e^{-\ii\EC(x)t}.\label{eq:Gamma-x}
\end{equation}
\Green{We emphasize that $x$ and $\nu(x)$ in Eq.~(\ref{eq:Gamma-x}) have their own meaning depending on the specific context of how the JJA bath is built. We will come back their exact meaning when we discuss specific examples in Sec.~\ref{sec:Examples}. Here, one can simply view $x$ as the index for the modes of the JJA and $\nu(x)$ is the degeneracy of the mode of $\EC(x)$}. Furthermore, we note that Eq.~(\ref{eq:Gamma-x}) is valid if the continuous versions
of Eqs.~(\ref{eq:EJ-small}) and~(\ref{eq:LowT}) hold. Namely,
for all $x$, we must have 
\begin{align}
 & \EJ(x)\ll\EC(x),\,\sqrt{\EC(x)\delta\EC},\label{eq:small-EJ-cont-limit}\\
 & \beta^{-1}\ll\Delta_{*}\equiv\min_{x}\Delta(x),\label{eq:zeroT-limit}
\end{align}
where $\delta\EC$ is the width of the distribution of the charging
energy across the chain and 
\begin{equation}
\Delta(x)\equiv\EC(x)/[-\ln\lambda(x)].\label{eq:TC-def}
\end{equation}
We deem the junction at position $x$ to be in \textit{the zero-temperature
limit} if $\beta^{-1}\ll\Delta(x)$. Then, the whole chain is in the
zero-temperature limit if all junctions are, that is if Eq.~(\ref{eq:zeroT-limit})
is satisfied.

We change variables in Eq.~(\ref{eq:Gamma-x}) to obtain 
\begin{equation}
\Gamma(t)=\left(\frac{\varepsilon_{\mathrm{I}}}{2e}\right)^{2}\int_{0}^{\infty}\dd{\EC}J(\EC)\e^{-\ii\EC t},\label{eq:Gamma}
\end{equation}
where 
\begin{equation}
J(\EC)\equiv2\sum_{k}\nu_{k}(\EC)[\EJ^{(k)}(\EC)]^{2}\abs{\dv{x_{k}}{\EC}\strut(\EC)}.\label{eq:JEC-def}
\end{equation}
Here $k$ labels the intervals~$I_{k}$ on which $\EC(x)$ is monotonic
so that the inverse function $x_{k}(\EC)$ is properly defined, and
\begin{align}
 & \nu_{k}(\EC)\equiv\nu(x_{k}(\EC)),\\
 & \EJ^{(k)}(\EC)\equiv\EJ(x_{k}(\EC)).
\end{align}
Comparing to Eq.~(\ref{eq:Gamma-HC}), we notice that $J(\EC)$ plays
a similar role to that of the spectral density in the harmonic regime,
\green{accounting for the} decay of the correlation function. Roughly
speaking, $\Gamma(t)$ will decay on a timescale of the order of
the width of $J(\EC)$ since $\Gamma(t)$ is the ``half Fourier transform''
of $J(\EC)$ as seen in Eq.~(\ref{eq:Gamma}). Then, if we denote
by $\delta\EC$ the width of $J(\EC)$, the characteristic decay time
of $\Gamma(t)$ will approximately be $1/\delta\EC$. 

\section{Temperature-driven transition to non-Markovianity \label{sec:non-Markovian}}

When temperature is much smaller than the charging energy but violates
Eq.~(\ref{eq:LowT}), the single-junction correlation function is
given by Eq.~(\ref{G_eps}) rather than Eq.~(\ref{G_lowT}). This
is because the next-to-leading order term $e^{-\beta\EC}$ becomes
comparable to the leading-order term. In the continuum limit, this
term contributes to a constant offset to the correlation function
when the bath temperature reaches the point $\beta^{-1}\sim\Delta_{*}$,
but $\beta^{-1}\ll\min_{x}\EC(x)$. 
\Green{ Thus, the Josephson bath correlation function no longer decays in this regime, meaning that the dynamics of the system becomes non-Markovian. The presence of this constant offset can be intuitively understood considering that in the regime we analyze, charge fluctuations in the Josephson chain are long-lived due to the small value of the Josephson energy (the commutator of the charge operator $N_\alpha$ with the chain Hamiltonian is very small). When the constant offset is nonnegligible, one can expect that correlations between the state of the system and the bath built up by the interaction never completely decay, which invalidates the GKSL treatment. 

To understand when this happens, we use Eqs.~(\ref{eq:g-alpha}) and~(\ref{eq:Gamma-t}),
to} calculate the constant offset, 
\begin{equation}
\Gamma_{0}=2\left(\frac{\varepsilon_{\mathrm{I}}}{e}\right)^{2}\int\mathrm{d}x\,\nu(x)\EC^{2}(x)e^{-\beta\EC(x)}.\label{eq:Gamma0}
\end{equation}
The magnitude of this offset will depend on the spatial
variation of $\Delta(x)$: We denote the place where $\Delta(x)$
reaches its minimum $\Delta_{*}$ as $x_{*}$. When the temperature
of the chain $\beta^{-1}$ becomes comparable to $\Delta_{*}$, obviously
the offset will contribute to the single-junction correlation function
for junctions near $x=x_{*}$. If the spatial variation of the $\Delta(x)$
is small, i.e., $\max_{x}\big|\Delta(x)-\Delta_{*}\big|$ is not too
much larger than $\Delta_{*}$, then the offset will contribute to a significant
portion of the single-junction correlation functions across the
chain, not just for junctions near $x=x_{*}$ so that the overall
bath correlation functions is significantly shifted.

However, if the spatial variation of $\Delta(x)$ is relatively
large, then it may happen that only a small portion of the chain near $x=x_{*}$
surpasses the zero-temperature limit when $\beta^{-1}$ becomes comparable
to $\Delta_{*}$. When this is the case, the offset of the bath correlation
function may be small compared to the magnitude of the one in the
zero-temperature limit, or even negligible.

We will discuss this phenomenon again for a concrete example with
effective Lorentz spectral density in Sec.~\ref{subsec:Example:-Lorentzian-spectral}.

\section{\label{sec:DecayLS}The GKSL master equation and bath duality}

\subsection{The decay rate and Lamb shift}

\pur{We now discuss the dynamics of the LC oscillator that is weakly
and capacitively coupled to the Josephson bath, shown in the dotted
box in Fig.~\ref{fig:setup}(a).} When the correlation function
$\Gamma(t)$ decays fast, the system dynamics may be described by
a GKSL master equation. Combined with the results of previous section,
it requires $\delta\EC$ be large enough, which can come from reservoir
engineering or disorder of the charging energy (see Sec.~\ref{sec:Examples}).
When this is the case, we can perform the Born--Markov approximation
and the secular approximation,\citep{CohenAtomPhoton,breuer2007thetheory}
to find the following (interaction picture) Markovian master equation
\pur{for the LC oscillator shown in the dotted box in Fig.~\ref{fig:setup}(a)}:
\begin{equation}
\frac{\mathrm{d}\rho_{\mathrm{S}}}{\mathrm{d}t}=-\ii[H_{\mathrm{LS}},\rho_{\mathrm{S}}(t)]+\kappa(\omega_{0})\mathcal{D}[b]\rho_{\mathrm{S}}(t).\label{eq:GKSL}
\end{equation}
Here \pur{$\omega_{0}$ is defined in Eq.~(\ref{eq:omega0-def}),}
$H_{\mathrm{LS}}=\delta_{\mathrm{LS}}(\omega_{0})b^{\dagger}b$ is
the Lamb shift Hamiltonian that accounts for the renormalization of
the LC oscillator's energy levels due to the coupling to the JJA bath,
while the dissipator~$\mathcal{D}[b]\rho_{\mathrm{S}}(t)\equiv b^{\dagger}\rho_{\mathrm{S}}(t)b-\{\rho_{\mathrm{S}}(t),b^{\dagger}b/2\}$
describes the equilibration of the oscillator with the bath through
the emission of photons at frequency~$\omega_{0}$. The Lamb shift~$\delta_{\mathrm{LS}}(\omega_{0})$
and the emission rate~$\kappa(\omega_{0})$ are both deduced from
the Fourier transformed correlation function~$\Gamma(\omega_{0})$,
\begin{align}
 & \delta_{\mathrm{LS}}(\omega_{0})=-\frac{\mathcal{C}\omega_{0}}{2}\Im\qty(\Gamma(\omega_{0})+\Gamma(-\omega_{0})),\label{eq:coeff-delta-omg}\\
 & \kappa(\omega_{0})=\mathcal{C}\omega_{0}\Re\Gamma(\omega_{0}),\label{eq:coeff-kappa-omg}\\
 & \Gamma(\omega_{0})=\int_{0}^{\infty}\dd{t}\Gamma(t)\e^{\ii\omega_{0}t}.\label{eq:coeff-Gamma-omg}
\end{align}
A brief derivation of the GKSL master equation is
presented in App.~\ref{sec:GKSL}.

Let us now calculate the Lamb
shift and decay rate stemming from the JJA bath correlation function
given in Eq.~(\ref{eq:Gamma}). The half Fourier transform yields
\begin{equation}
\Gamma(\omega)=\left(\frac{\varepsilon_{\mathrm{I}}}{2e}\right)^{2}\left[\pi J(\omega)\Theta(\omega)-\ii\mathcal{P}\negthickspace\int_{0}^{\infty}\dd{\EC}\frac{J(\EC)}{\EC-\omega}\right],\label{eq:gamma-omega}
\end{equation}
where $\Theta(\omega)$ is the Heaviside function and $\mathcal{P}$
denotes the Cauchy principal value. We then straightforwardly obtain
the Lamb shift, 
\begin{equation}
\delta_{\mathrm{LS}}(\omega_{0})=\frac{\omega_{0}\varepsilon_{\mathrm{I}}^{2}}{16E_{\mathrm{Q}}}\mathcal{P}\negthickspace\int_{0}^{\infty}\dd{\EC}J(\EC)\left(\frac{1}{\EC-\omega_{0}}+\frac{1}{\EC+\omega_{0}}\right),\label{eq:Lamb-shift}
\end{equation}
where we have introduced $E_{\mathrm{Q}}=e^{2}/(2\mathcal{C})$, the
renormalized charging energy of the LC oscillator. Finally, the decay
rate reads 
\begin{equation}
\kappa(\omega_{0})=\frac{\pi\varepsilon_{\mathrm{I}}^{2}\omega_{0}J(\omega_{0})}{8E_{\mathrm{Q}}},\label{eq:decay-rate}
\end{equation}
which is the rate of emission of a photon at frequency~$\omega_{0}$
by the LC oscillator. The rate for the opposite process, where the
oscillator is excited by the absorption of a photon vanishes here.
This indicates that the JJA bath is effectively at zero temperature:
The oscillator cannot absorb any photon if the bath emits none.


\subsection{Bath duality\label{subsec:Mapping}}

From Eqs.~(\ref{eq:Gamma-EJ-0T}) and~(\ref{eq:Gamma}), it is clear
that, the zero-temperature correlation function of the JJA bath takes
similar forms whether $\EJ$ or $\EC$ defines the largest energy
scale in the problem, leading to similar effects on the system's dynamics.
The only difference between the two regimes is then how the spectral
density, and therefore the damping rate and Lamb shift, are related
to the microscopic parameters of the JJA.

We found that the spectral densities in the two regimes can be mapped
onto each other, e.g., through the parameter correspondence list in
Tab.~\ref{tab:Parameter-correspondence}. 
The existence of such mapping implies that there exist two sets of
parameters, ($\EJ(x),\,\EC(x)$) in the large $\EC$ regime and ($\tilde{E}_{\mathrm{J}}(x),\tilde{E}_{\CC}(x)$)
in the large $\EJ$ regime, linked by the relation in Table~\ref{tab:Parameter-correspondence},
which lead to the same coarse-grained dynamics for any small system
coupled to the JJA, i.e., the same form of GKSL master equation and
the same value of the coefficients. We dub the mapping the between
the two regimes as \textit{bath duality}.

Let us now illustrate how the bath duality is actually performed.
The parameters in the large $\EJ$ regime come with tildes to
distinguish them from those in the large $\EC$ regime. Moreover,
the parameter $\EC(x_{k})$ in the left column plays the same role
as the frequency $\tilde{\omega}_{k}$ in the right column.

To obtain the same dynamics, one can ensure that the value
of $\EC(x_{k})$ (second row of the table) is the same as that of
$\tilde{\omega}(x_{k})$, that the function $\nu_{k}(\EC)=\nu(x_{k}(\EC))$
(third row) takes the same values as $\tilde{\nu}_{k}(\tilde{\omega})=\tilde{\nu}(x_{k}(\tilde{\omega}))$,
and finally that $2\{\EJ^{(k)}(\EC)\}^{2}/\EC=2\{\EJ(x_{k}(\EC))\}^{2}/x_{k}(\EC)$
(last row of the table) equals $\tilde{E}_{\CC}^{(k)}(\tilde{\omega})=\tilde{E}_{\CC}(x_{k}(\omega))$.

From this correspondence rules, we find 
\begin{equation}
\tilde{E}_{\CC}(x_{k})=\frac{2\EJ^{2}(x_{k})}{\EC(x_{k})},\label{eq:EC-tilde-dist}
\end{equation}
and 
\begin{equation}
\tilde{E}_{\mathrm{J}}(x_{k})=\frac{\tilde{\omega}^{2}(x_{k})}{2\tilde{E}_{\CC}(x_{k})}=\frac{\tilde{\omega}^{3}(x_{k})}{4\EJ^{2}(x_{k})}=\frac{\EC^{3}(x_{k})}{4\EJ^{2}(x_{k})},\label{eq:EJ-tilde-dist}
\end{equation}
using the identification of $\tilde{\omega}(x_{k})$ with $\EC(x_{k})$.
This leads to the relation 
\begin{equation}
\tilde{E}_{\CC}(x_{k})/\tilde{E}_{\mathrm{J}}(x_{k})\sim[\EJ(x_{k})/\EC(x_{k})]^{4}\ll1,
\end{equation}
which shows the mapping indeed connects the large $\EC$ regime to
the large $\EJ$ regime. Furthermore, one can also check that the
mapping preserves the zero-temperature limit 
\begin{equation}
\tilde{\beta}\ll\tilde{\omega}(x_{k}),\,\forall x_{k}\in I_{k}
\end{equation}
as along as the temperature for the large $\EJ$ regime is taken as
\begin{equation}
\tilde{\beta}^{-1}\lesssim\ln\left[\EC(x)/\EJ(x)\right]_{\mathrm{min}}\beta^{-1}.
\end{equation}

\begin{table}
\begin{centering}
\begin{tabular}{c|c}
Large $\EC$ limit & Large $\EJ$ limit\tabularnewline
$\EJ(x)\ll\EC(x)$  & $\tilde{E}_{\CC}(x)\ll\tilde{E}_{\mathrm{J}}(x)$\tabularnewline
\hline
Zero-temperature limit &  Zero-temperature limit
\tabularnewline
$\beta^{-1}\ll \Delta_{*}$  & $\tilde{\beta}^{-1}\ll\min_{x}\tilde{\omega}(x)$\tabularnewline
\hline 
$x_{k}$$\in I_{k}$  & $x_{k}\in I_{k}$\tabularnewline
\hline 
$\EC(x_{k})$  & $\tilde{\omega}(x_{k})\equiv\sqrt{2\tilde{E}_{\CC}(x_{k})\tilde{E}_{\mathrm{J}}(x_{k})}$\tabularnewline
\hline 
$\nu_{k}(\EC)$  & $\tilde{\nu}_{k}(\tilde{\omega})$\tabularnewline
\hline 
$2[\EJ^{(k)}(\EC)]^{2}$/$\EC$  & $\tilde{E}_{\CC}^{(k)}(\tilde{\omega})$\tabularnewline
\end{tabular}
\par\end{centering}
\caption{\label{tab:Parameter-correspondence} Parameter correspondence between
the large $\EC$ regime and large $\EJ$ regime at the zero temperature
limit, where the leading order correlation function of JJA bath produces
the same coarse-grained dynamics for the primary LC oscillator. The
parameters in the large $\EJ$ regime come with tildes to distinguish
those in the large $\EC$ regime. The basic variable on the left column
is $\EC$ while the one on the right column is $\tilde{\omega}$.}
\end{table}

An example of how this mapping can be implemented will be discussed
in next section.

\section{Examples\label{sec:Examples}}

In this section, we will give two concrete examples illustrating how
the result in Sec.~\ref{sec:DecayLS} can be applied. Practically,
we know\citep{orlando1991foundations}
\begin{align}
E_{\CC\alpha} & =\frac{2e^{2}w_{\alpha}}{\epsilon_\mathrm rA_{\alpha}},\label{eq:EC-experiment}\\
E_{\JJ\alpha} & =\frac{F_{\JJ}A_{\alpha}}{\zeta\sinh(w_{\alpha}/\zeta)},\label{eq:EJ-experiment}
\end{align}
where $w_{\alpha}$ and $\epsilon_\mathrm r$ are the thickness and dielectric
constant of the oxidation material, $A_{\alpha}$ is the area of the
junctions, $F_\JJ$ is a constant that depends on the density
of the cooper pairs on the islands which has the dimension of a force,
and $\zeta$ is a constant that depends on the tunneling barrier that
has the dimension of a length. \blue{Below, we will consider variations of the oxidation material thickness and of the junction area across the chain, all other parameters being kept fixed}. 

In the first example, the junction density and the junction area distribution are properly engineered such that the effective
spectral density is Lorentzian, a case frequently studied in the field
of open quantum systems. In the second example, we discuss a common
case in condensed matter physics where the JJA presents disorder in the
values of the charging and Josephson energies caused by the Gaussian
disorder of the width of the oxidation material. 
It should be noted that in the first example the spatial dependences
of the charging and Josephson energies on the junction position are
known, while in the second example these energies at each junction
are random and correlated through the width of the oxidation material.

\begin{figure}
\centering
\includegraphics[width=\linewidth]{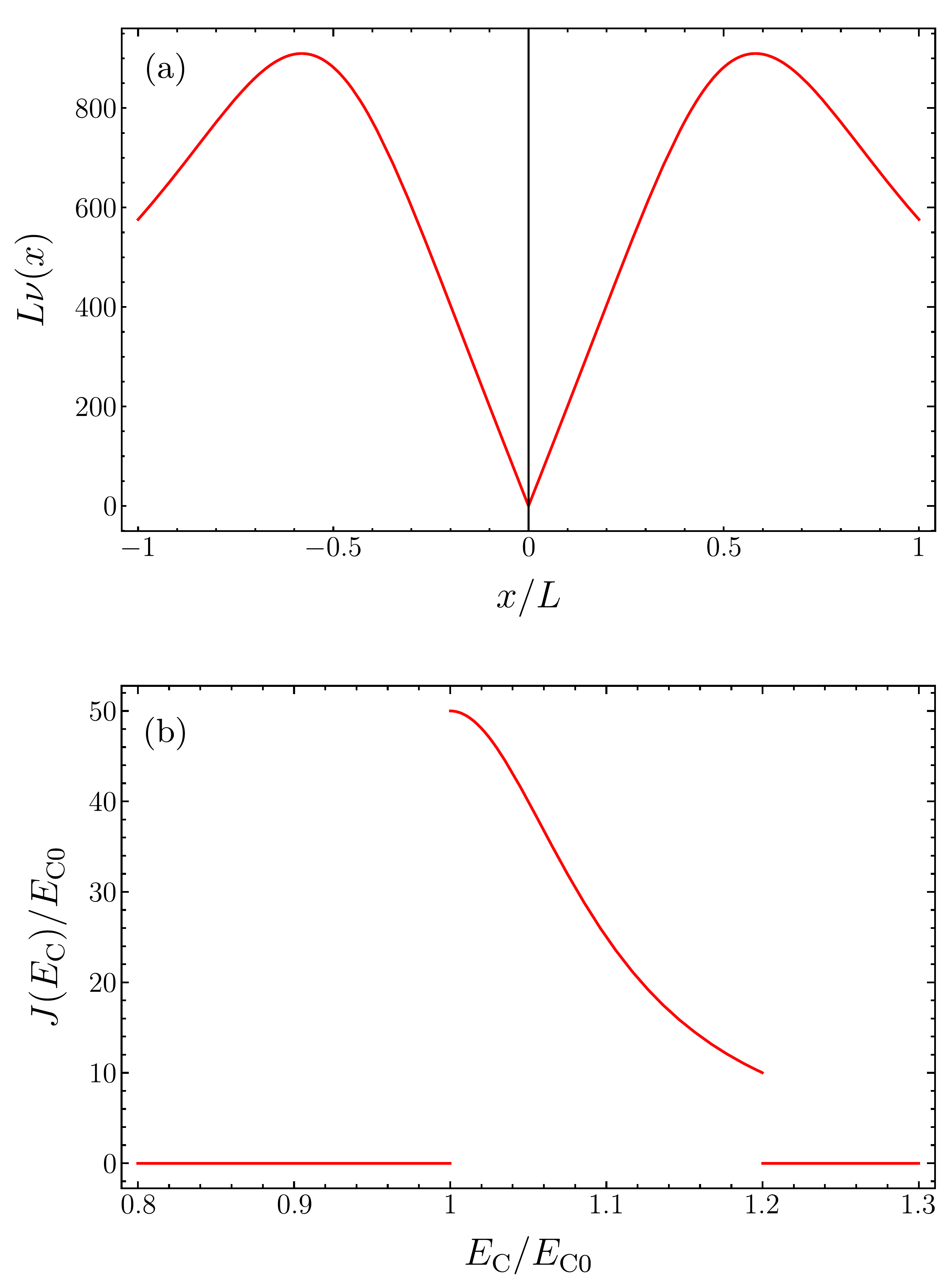}
\caption{\label{fig:junction-property}(a) The junction density given by Eq.~(\ref{eq:Junction-density})
(b) the spectral density given by Eq.~(\ref{eq:Spectral density}),
\red{which is nonvanishing only in the range of characteristic frequencies
of the Josephson bath, i.e., $[E_{\CC0},\,(1+a/2)E_{\CC0}${]}}.
Value of parameters: $\mathcal{A}=500$, $\sigma=0.25$, $a=0.4$,
$E_{\JJ0}=0.05E_{\CC0}$. The total number of the junction
is the JJA is roughly $N_{\mathrm{J}}=\int_{-L}^{L}\mathrm dx\,\nu(x)\approx1277$.}
\end{figure}

\subsection{Engineering the areas of the junctions\label{subsec:Example:-Lorentzian-spectral}}

\blue{We assume all parameters are kept uniform along the chain, except the area of the junctions which is engineered such
that}
\begin{equation}
\EC(x)=\left(1+\frac{ax^{2}}{2L^{2}}\right)E_{\CC0},\label{eq:EC-quadratic}
\end{equation}
where $a>0$ is dimensionless quantity characterizing the variation
of the charging energy across the chain, $x\in[-L,L]$. According
to Eqs.~(\ref{eq:EC-experiment}) and~(\ref{eq:EJ-experiment}), the distribution
of $\EJ$ across the chain is
\begin{equation}
\EJ(x)=E_{\JJ0}\left(1+\frac{ax^{2}}{2L^{2}}\right)^{-1},
\end{equation}
where 
\begin{equation}
E_{\JJ0}=\frac{2e^{2}F_{\JJ}w}{\epsilon_\mathrm r\zeta E_{\CC0}\sinh(w/\zeta)}.
\end{equation}
To satisfy the large $E_{\CC}$ condition in Eq.~(\ref{eq:small-EJ-cont-limit}),
we require $E_{\CC0}\gg E_{\JJ0}$. The junctions are assumed
to be distributed according to the junction density

\begin{equation}
\nu(x)=\frac{\mathcal{A}L^{2}\big|x\big|}{x^{4}+4\sigma^{2}L^{4}}\left(1+\frac{ax^{2}}{2L^{2}}\right)^{2},\label{eq:Junction-density}
\end{equation}
where the dimensionless quantities $\mathcal{A}$ and $\sigma$ \green{respectively
characterize the amplitude and variation} of the junction density
across the chain. \Green{The junction density $\nu(x)$ accounts for the weights of a junction with parameter $\EC(x)$ and $\EJ(x)$, similarly to the concept of density of states accounting for the degeneracy of energy levels.} The shape of $\nu(x)$ is shown in Fig.~\ref{fig:junction-property}(a).
Then, according to Eq.~(\ref{eq:JEC-def}), 
\begin{equation}
J(\EC)=\frac{a\mathcal{A}E_{\CC0}E_{\JJ0}^{2}}{(\EC-E_{\CC0})^{2}+(a\sigma E_{\CC0})^{2}},\label{eq:Spectral density}
\end{equation}
when $E_{\CC0}\le\EC\le(1+a/2)E_{\CC0}$, and $J(\EC)=0$
otherwise. The plot of $J(\EC)$ is shown in Fig.~\ref{fig:junction-property}(b).


\begin{figure}
\centering
\includegraphics[width=\linewidth]{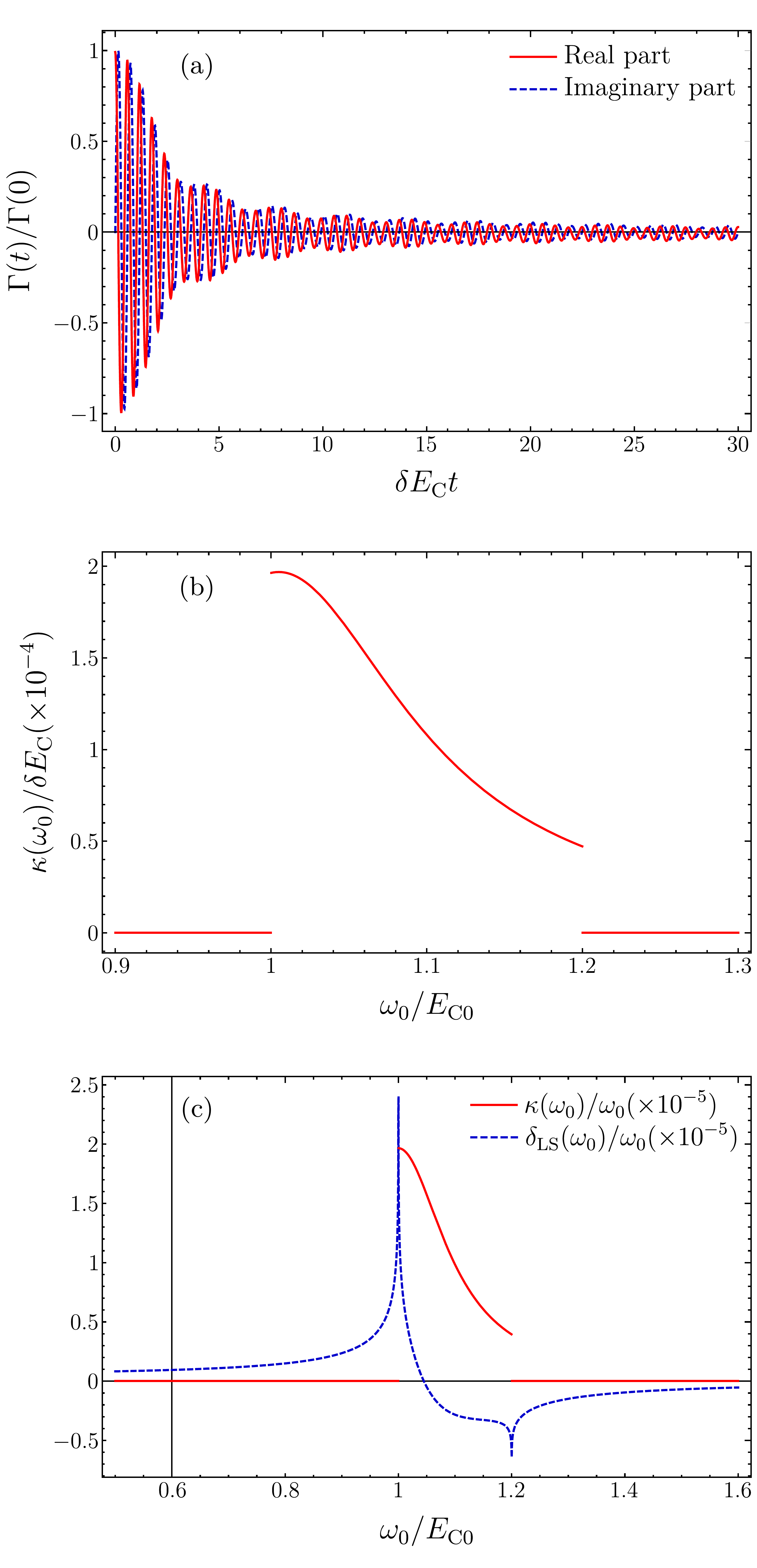}
\caption{\Green{(a) The normalized correlation function of the JJA, $\Gamma(t)/\Re[\Gamma(0)]$.
(b) The decay rate (measured in units of $\delta\EC$) versus the frequency of the LC oscillator (measured in units of $E_{\mathrm C0}$). (c)
The Lamb shift and decay rate (measured in units of $\omega_0$) versus the frequency of the LC oscillator (measured in units of $E_{\mathrm C0}$).} The chain is in the limit of zero-temperature given
by Eq.~(\ref{eq:zeroT-limit}), where the bath correlation function
decays rapidly compared to the decay timescale of the oscillator,
as shown in (a) and (b). Other parameters for the junction chain is
the same as Fig.~\ref{fig:junction-property}. The value for the
LC oscillator and the coupling $\varepsilon_{\mathrm{I}}=0.01$, $E_{\mathrm{Q}}=100E_{\CC0}$
and $\delta\EC=0.1E_{\CC0}$. \red{The decay rate, proportional
to the spectral density shown in Fig.~\ref{fig:junction-property},
is nonzero only when the frequency of the LC oscillator $\omega_{0}$
is resonant with the characteristic modes in the Josephson bath, whose
frequencies lie in the range $[E_{\CC0},\,(1+a/2)E_{\CC0}]$}.
Comparing Figs.~(a) and~(b), one obviously observe that the decay of
the bath correlation function is much fast than the decay of the LC
oscillator so that the Born--Markov approximation is satisfied in
this case. The red line in Fig.~(c) indicates that the decay rate
is much smaller than the frequency of the LC oscillator such that
the secular approximation is satisfied. \label{fig:property-LC}}
\end{figure}

To justify the Born--Markov and secular approximations,
the following conditions must be satisfied 
\begin{equation}
\kappa(\omega_{0})\ll\omega_{\mathrm{B}},\omega_{0},\label{eq:BM-secular}
\end{equation}
where $\omega_{\mathrm{B}}^{-1}$ is the timescale, on which the
JJA bath correlation function~$\Gamma(t)$ decays.

Now we derive an \textit{empirical} rule for the above example, under
which Eq.~(\ref{eq:BM-secular}) can be satisfied. Similar rules
can be analogously derived for other distributions of the junction
density and parameters. According to Eq.~(\ref{eq:Gamma}), one can
roughly regard $\Gamma(t)$ as the Fourier transform of $J(\EC)$,
although, strictly speaking, it is ``the half Fourier transform''
$J(\EC)$. Therefore the decay rate of $\Gamma(t)$ should be the
width of $J(\EC)$, i.e., 
\begin{equation}
\omega_{\mathrm{B}}\sim\delta\EC\equiv aE_{\CC0}\min\left\{\sigma,\frac12\right\}.\label{eq:omega-B}
\end{equation}
According to Eqs.~(\ref{eq:decay-rate}) and~(\ref{eq:Spectral density}),
the maximum decay rate is obtained for $\omega_{0}=E_{\CC0}\sqrt{1+(a\sigma)^{2}}$,
with $\kappa_{\mathrm{max}}=\zeta_{\mathrm{M}}E_{\CC0}\left(1+\sqrt{1+(a\sigma)^{2}}\right)/2$,
where 
\begin{equation}
\zeta_{\mathrm{M}}\equiv\frac{\pi\mathcal{A}\varepsilon_{\mathrm{I}}^{2}E_{\JJ0}^{2}}{8a\sigma^{2}E_{\CC0}E_{\mathrm{Q}}}
\end{equation}
characterizes the Markovianity of the chain. The maximum ratio between
$\kappa(\omega_{0})$ and $\omega_{0}$ occurs when $\omega_{0}=E_{\CC0}$,
where $[\kappa(\omega_{0})/\omega_{0}]_{\mathrm{max}}=\zeta_{\mathrm{M}}$.
Therefore, Eq.~(\ref{eq:BM-secular}) can be satisfied if
\begin{align}
 & \frac{\zeta_{\mathrm{M}}[1+\sqrt{1+(a\sigma)^{2}}]}{a\sigma}\ll1,\,2\sigma,\label{eq:empirical-BM}\\
 & \zeta_{\mathrm{M}}\ll1.\label{eq:empirical-Secular}
\end{align}
It turns out that Eqs.~(\ref{eq:empirical-BM}) and~(\ref{eq:empirical-Secular})
are satisfied when $\zeta_{\mathrm{M}}$ is small enough.
In such a case, the JJA behaves as a Markovian bath. In Fig.~\ref{fig:property-LC},
one can see that for the given parameters for the JJA and the LC oscillator,
Eqs.~(\ref{eq:empirical-BM}) and~(\ref{eq:empirical-Secular})
are satisfied so that the validity of the GKSL master equation is
justified. \pur{From Figs.~\ref{fig:property-LC}(a) and~(b), we
observe that the Josephson bath correlation time $\omega_{\mathrm{B}}^{-1}$
is roughly about $10/\delta\EC$ and $\kappa/\omega_{\mathrm{B}}\sim10^{-4}\text{--}10^{-3}$.
Thus within the GKSL weak-coupling approach, the coupling that the
circuit QED setup here has achieved is already much larger compared
to the atomic case where\citep{breuer2007thetheory,CohenAtomPhoton} $\kappa/\omega_{\mathrm{B}}\sim10^{-7}\text{--}10^{-6}$.
}

\begin{figure}
\centering
\includegraphics[width=\linewidth]{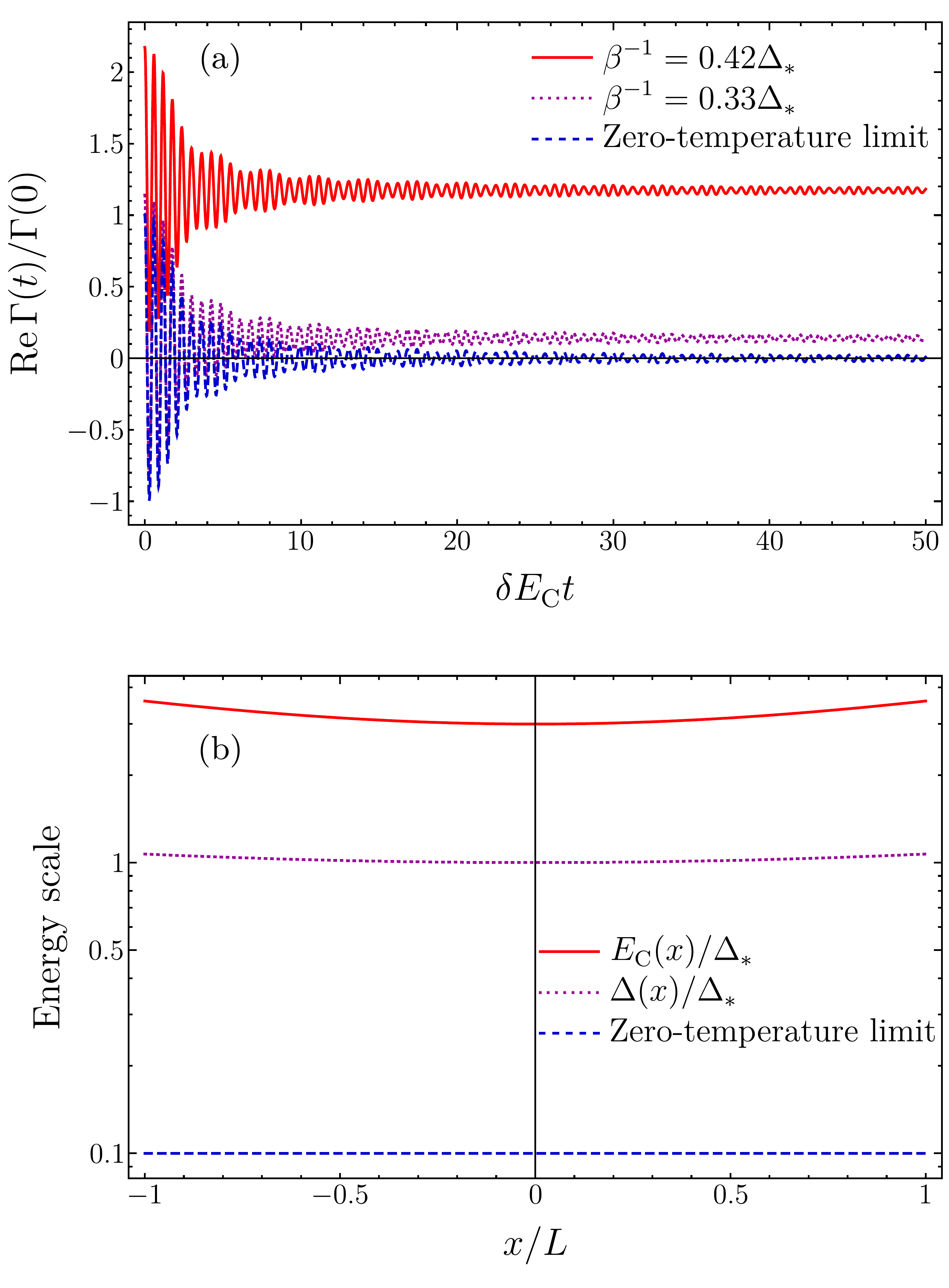}
\caption{\label{fig:smallWidthEC}\Green{(a) The real part of the normalized
correlation function versus time. (b) The distribution of the energy
scales of the $\Delta(x)$, $\EC(x)$ and the zero-temperature limit
(given by the blue dashed lines) are $\beta^{-1}\lesssim\Delta_{*}/10$.}
The junction parameters are the same as described in the caption of Fig.~\ref{fig:junction-property}.
In particular, $\EC\in[1,\,1.2]E_{\CC0}$, $E_{\JJ0}=E_{\CC0}/20$,
$\delta E_{\CC}=0.1E_{\CC0}$, and $\Delta_{*}=\Delta(0)=0.33E_{\CC0}$.
The normalization is performed by dividing the value of the real part
of correlation function in the zero-temperature limit evaluated at
$t=0$. The blue dashed line in (a) corresponding to the normalized
correlation function in the zero-temperature limit is the same as the
one shown in Fig.~\ref{fig:property-LC}. The offset in the correlation
function for $\beta^{-1}=0.42\Delta_{*}$ shown in red solid line
in (a) is about $100\%$ when compared to the magnitude for the correlation
function in the zero-temperature limit. We note that the shape of
$\Delta(x)$ in (b) is quite flat. This indicates when the temperature
is increased beyond the zero-temperature limit for $x_{*}=0$, significant
portion of near the origin will also violate the zero-temperature.
This accounts for the huge offset in (a). }
\end{figure}

\begin{figure}
\centering
\includegraphics[width=\linewidth]{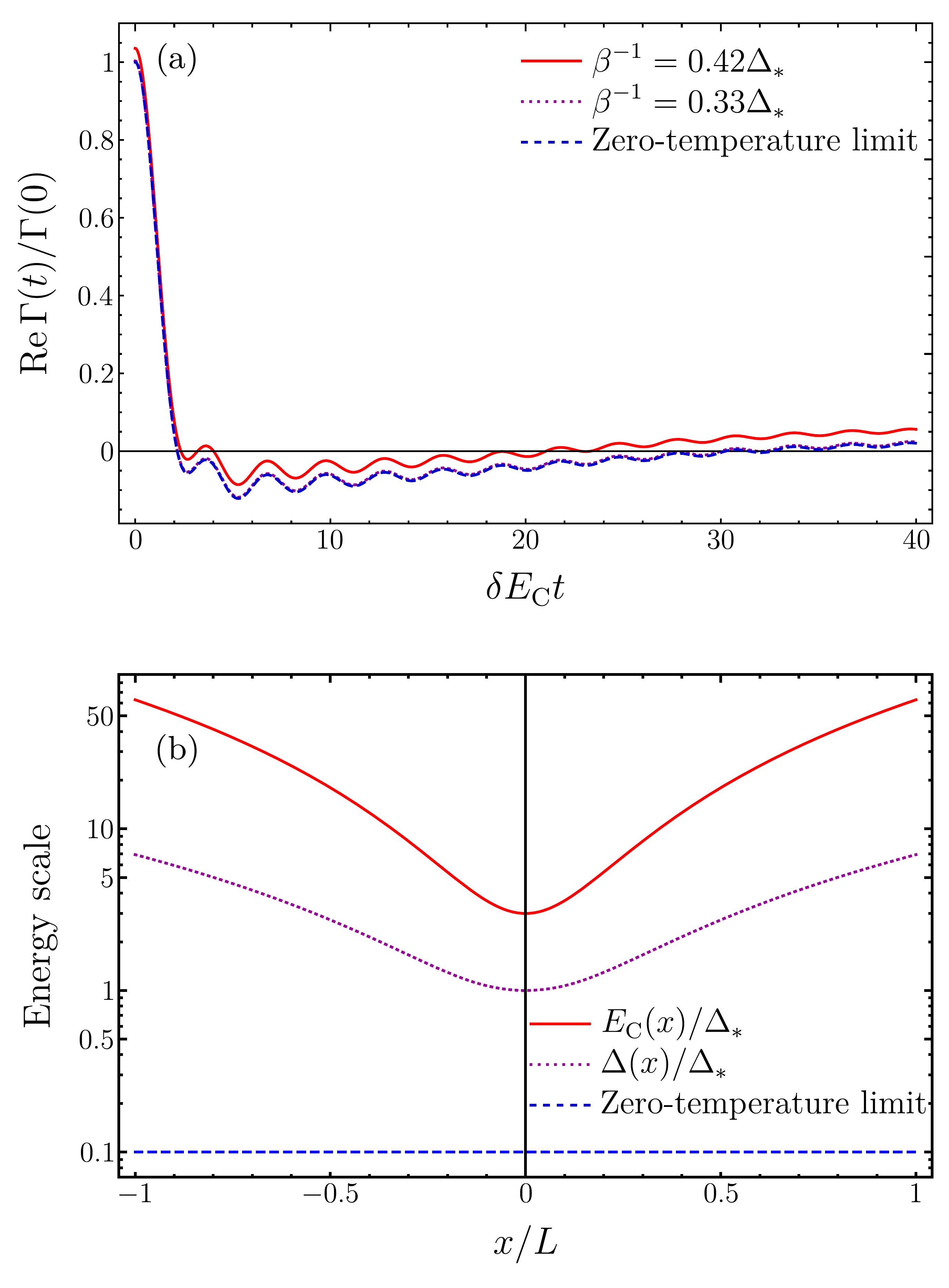} 
\caption{\label{fig:largeWidth}\Green{(a) The real part of the normalized
correlation function versus time. (b) The distribution of the energy
scales of the $\Delta(x)$, $\EC(x)$ and the zero-temperature limit
(given by the blue dashed lines) are $\beta^{-1}\lesssim\Delta_{*}/10$.}
The junction parameter $a=40$ so that $\EC\in[1,\,21]E_{0}$, wider
than Fig.~\ref{fig:smallWidthEC}. Other parameters: $E_{\JJ0}=E_{\CC0}/20$,
$\mathcal{A}=50$, $\sigma=0.25$, $E_{\mathrm{Q}}=100E_{\CC0}$,
$\varepsilon_{\mathrm{I}}=0.01$, $\delta\EC=10E_{\CC0}$ and
$\Delta_{*}=\Delta(0)=0.33E_{\CC0}$. The normalization of the
correlation function in (a) is performed by dividing the value of
the real part of the correlation function in the zero-temperature
limit evaluated at $t=0$. The blue dashed line in (a) corresponding
to the normalized correlation function in the zero-temperature limit is
the same as the one shown in Fig.~\ref{fig:property-LC}. The offset
in the correlation function for $\beta^{-1}=0.42\Delta_{*}$ shown
in red solid line in (a) is about $5\%$ when compared to the magnitude
for the correlation function in the zero-temperature limit. Note that
the shape of $\Delta(x)$ in (b) is more warped than that in Fig.~\ref{fig:smallWidthEC}(b).
Therefore, we expect when the temperature is increased beyond the
zero-temperature limit near $x_{*}=0$, offset in the correlation
function should be smaller than that in Fig.~\ref{fig:smallWidthEC}(a).}
\end{figure}

Note that the Markovian property of the Josephson bath, where the
correlation function decays faster than the LC oscillator, as shown
in Fig.~\ref{fig:property-LC}(a), only holds in the zero-temperature
limit given by Eq.~(\ref{eq:zeroT-limit}). As we have discussed
in Sec.~\ref{sec:non-Markovian}, when the temperature of the chain
is comparable to $\Delta_{*}$ defined in Eq.~(\ref{eq:zeroT-limit}),
non-Markovian dynamics may occur due to the significant constant offset
in the Josephson bath correlation function. For the charging energy
distribution in Eq.~(\ref{eq:EC-quadratic}), $\Delta_{*}$ is $E_{\CC0}/\ln(E_{\CC0}/E_{\JJ0})$,
which is reached at the origin. The magnitude of the offset depends
on the spatial variation of $\Delta(x)$ defined in Eq.~(\ref{eq:TC-def}):
The flatter the spatial distribution, the larger the offset. The correlation
function for cases with small and large spatial variation of $\Delta(x)$
are given in Figs.~\ref{fig:smallWidthEC} and~\ref{fig:largeWidth},
respectively. One can easily find that for $\beta^{-1}=0.33\Delta_{*}$
and $\beta^{-1}=0.42\Delta_{*}$, the portion in the JJA near the
origin in Fig.~\ref{fig:smallWidthEC}(b) that surpasses the zero-temperature
limit is larger than the one in Fig.~\ref{fig:largeWidth}(b). This
is why the offset in Fig.~\ref{fig:smallWidthEC}(a) is larger when
compared to Fig.~\ref{fig:largeWidth}(a).

\blue{Finally, we demonstrate how the parameter correspondence discussed in Sec.~\ref{subsec:Mapping}
can be implemented with the distributions in Eqs.~(\ref{eq:EC-quadratic})
and~(\ref{eq:Junction-density}). In the mapped large $\EJ$ regime,
the distribution of $\tilde{\omega}$ can be obtained by replacing
$\EC$ in Eq.~(\ref{eq:EC-quadratic}) with $\tilde{\omega}$ 
\begin{equation}
\tilde{\omega}(x_{k})=\left(1+\frac{ax_{k}^{2}}{2L^{2}}\right)E_{0},
\end{equation}
where $x_{1}\in[0,\,L]$ and $x_{2}\in[-L,\,0]$. According to Eqs.~(\ref{eq:EC-tilde-dist})
and~(\ref{eq:EJ-tilde-dist}), the distributions for $\tilde{E}_{\CC}$
and $\tilde{E}_{\mathrm{J}}$ are 
\begin{align}
 & \tilde{E}_{\CC}(x_{k})=\frac{2E_{\JJ0}^{2}}{E_{\CC0}[1+ax_{k}^{2}/(2L^{2})]^{3}},\\
 & \tilde{E}_{\mathrm{J}}(x_{k})=\frac{E_{\CC0}^{3}[1+ax_{k}^{2}/(2L^{2})]^{5}}{4E_{\JJ0}^{2}}.
\end{align}
These expressions define the distributions for the charging and Josephson
energies in the large $\EJ$ regime if we require the two regimes
give the same coarse-grained dynamics for the LC oscillator.}

\Green{ 

\subsection{Gaussian disorder in the oxidation thickness\label{subsec:disorder-EC}}

} 

In this section, \blue{we assume that the only source of the disorder of the charging
and Josephson energies is the random distribution of the
oxidation thickness in between two neighboring superconducting islands. All other properties for the junctions
in the chain are assumed to be the same, including the material of the oxidation,
the junction areas.}
We denote the oxidation thickness at junction as $w_{\alpha}$ and
assume it is a Gaussian random variable with mean $w_{0}$ and width
$\delta w$. From Eqs.~(\ref{eq:EC-experiment}) and~(\ref{eq:EJ-experiment}),
we immediately \blue{deduce that 
the charging and Josephson energies are correlated. To ensure that the condition in Eq.~(\ref{eq:small-EJ-cont-limit}) defining the large charging energy regime is satisfied, it is necessary that $E_{\CC\alpha}\gg E_{\JJ\alpha}$ for (almost) all the junctions in the chain. Since $w\sinh(w/\zeta)$ is a monotonically increasing function, the large charging energy regime is then guaranteed if
\begin{equation}
w_\mathrm{min}\sinh(\frac{w_{\mathrm{min}}}{\zeta})\gg\frac{\epsilon_\mathrm rF_\JJ A^2}{2e^2\zeta},\label{eq:Large-EC-disorder}
\end{equation}
where $w_\mathrm{min}$ is the minimum value of $w_\alpha$. Here, we assume that the probability  distribution for $w_\alpha$ for $w>w_\mathrm{min}$ is
\begin{equation}
\mathcal{P}(w)\equiv\sqrt{\frac{1}{2\pi\delta w^{2}}}\exp\left[-\frac{(w-w_{0})^{2}}{2\delta w^{2}}\right],
\end{equation}
while $\mathcal P(w)=0$ for $w<w_\mathrm{min}$.}

\begin{figure}





\centering
\includegraphics[width=\linewidth]{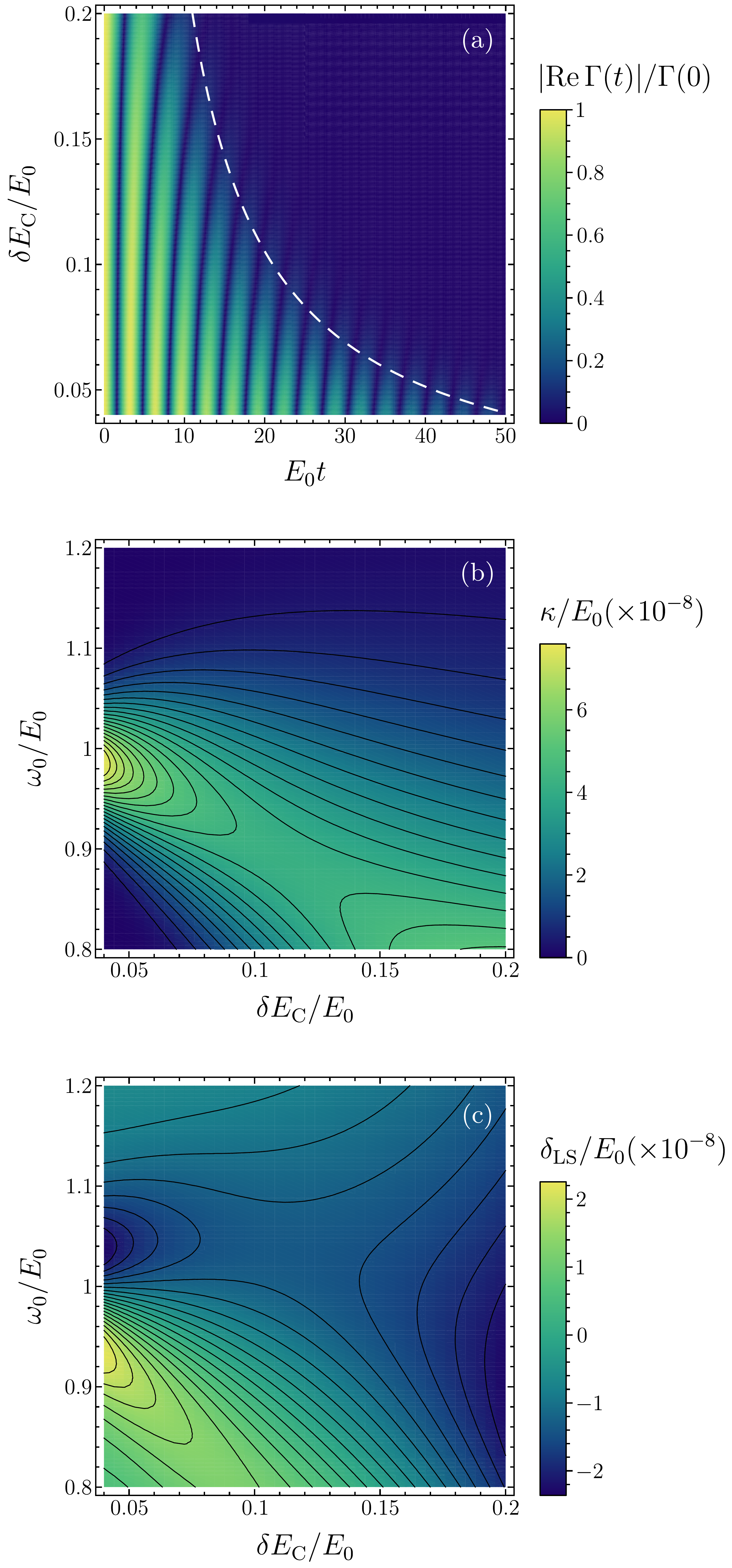}
\caption{\label{fig:2DPlots-Gaussian}2D plots of (a) the normalized magnitude
of the real part of the Josephson bath correlation function $\left|\Re\Gamma(t)\right|/\Gamma(0)$
versus time and the width of the disorder (b) the normalized decay
rate $\kappa/E_{0}$ versus the width of the disorder and the oscillator
frequency (c) the normalized Lamb shift $\delta_{\mathrm{LS}}/E_{0}$
versus the width of the disorder and the oscillator frequency. Values
of parameters: $F_{\JJ}A/\zeta=E_{0}/100$, $E_{\mathrm{min}}=E_{\zeta}=E_{0}/5$,
$E_{\mathrm{Q}}=2E_{0}$, $\varepsilon_{\mathrm{I}}=0.01$, and $N_{\JJ}=10000$.}
\end{figure}

In the continuum limit, the correlation function Eq.~(\ref{eq:Gamma-discrete})
can be written as 
\begin{equation}
\Gamma(t)=\frac{\varepsilon_\mathrm I^2N_{\JJ}F_{\JJ}^{2}A^2}{2e^2\zeta^2}\int_{w_{\mathrm{min}}}^{\infty}\mathrm dw\,\frac{\mathcal{P}(w)}{\sinh^{2}(w/\zeta)}\e^{2\ii e^2wt/(\epsilon_\mathrm rA)}.\label{eq:Gamma-disorder}
\end{equation}
\Green{Comparing with Eq.~(\ref{eq:Gamma-x}), we see that $w$ and $N_{\JJ}\mathcal{P}(w)$ play the roles of $x$ and $\nu(x)$ respectively.} \blue{Through an obvious change of variables in in Eq.~(\ref{eq:Gamma-disorder}), we obtain the following spectral density
\begin{equation}
    J(\EC)=\frac{2N_{\JJ}F_{\JJ}^{2}A^2}{\zeta^2\sinh^{2}(\EC/E_{\zeta})}\sqrt{\frac{1}{2\pi\delta\EC^{2}}}\exp\left[-\frac{(\EC-E_{0})^{2}}{2\delta\EC^{2}}\right],
\end{equation}
for $\EC>E_{\mathrm{min}}$, and $J(E_\CC)=0$ for $\EC<E_{\mathrm{min}}$, with
\begin{equation}
    \frac{E_{\mathrm{min}}}{w_\mathrm{min}}=\frac{E_0}{w_0}=\frac{\delta\EC}{\delta w}=\frac{E_\zeta}\zeta=\frac{2e^2}{\epsilon_\mathrm r A}
\end{equation}
Eq.~(\ref{eq:Large-EC-disorder}) now becomes $E_{\mathrm{min}}\sinh(E_{\mathrm{min}}/E_{\zeta})\gg F_\JJ A/\zeta$.}
The plots of correlation function, decay rate and Lamb shift are shown
in Fig.~\ref{fig:2DPlots-Gaussian}. While it is clear that the JJA
bath correlation time decreases with the width of the disorder $\delta\EC$,
it is straightforward to show that for fixed oscillator frequency
$\omega_{0}$ and $E_{0}$, the decay rate reaches its maximum when
$\delta\EC=\sqrt{2}|\omega_{0}-E_{0}|$. \pur{Note that in Fig.~\ref{fig:2DPlots-Gaussian}(b),
the ratio of $\kappa/\omega_\mathrm{B}\sim\kappa/(10\delta E_{\mathrm{C}})$ is of the order $10^{-5}$,
similar to the atomic case where\citep{breuer2007thetheory,CohenAtomPhoton} $\kappa/\omega_\mathrm{B}\sim10^{-7}\text{--}10^{-6}$.\blue{ Analogously to Sec.~\ref{subsec:Example:-Lorentzian-spectral}, one
can of course derive an empirical Markovianity criterion, and demonstrate
the non-Markovianity beyond the zero-temperature limit, which will
not be repeated here.}
}

\section{Discussion and Conclusion \label{sec:Discussions}}

We showed that when the charging energy is the largest energy scale
comparing to temperature and the Josephson energy, a 1D Josephson
junctions array (JJA) can behave as a Markovian bath \pur{of nonlinear
rotors} provided the distribution for the chain parameters meet specific
conditions, namely, the leading-order of the bath correlation function
decays rapidly. We calculated the dynamics of an LC oscillator that
is coupled to the JJA the using approach of the GKSL Markovian master
equation. In particular, we derived explicit expressions of the Lamb
shift and decay rate of the LC oscillator caused by the coupling to
the JJA. We found the leading order of the JJA bath correlation function
in the large charging energy regime bears the same form as that of
a harmonic bath, which can be approximated in the large Josephson
energy regime to the leading order. Based on this observation, we
established a mapping between the junction parameters in the two regimes,
identifying the set of parameters inducing the same coarse-grained
dynamics for a small quantum system coupled to the chain. We gave
two specific examples and showed that it fits into the GKSL framework.
In the first example the spatial distributions of junction density
and charging energy are properly engineered such that effective spectral
density is a Lorentzian while in the second example the charging energy
is a Gaussian random variable so that the effective spectral density
is a Gaussian.

When the temperature is increased to the point where a large region
across chain is beyond the zero-temperature limit, the JJA bath correlation
function gets significantly shifted by a constant, which renders the
dynamics of the LC oscillator non-Markovian. This phenomenon is an
indication that beyond the zero-temperature limit the primary system
is correlated with the JJA bath on a very long timescale, which cannot
be addressed within the framework of the GKSL master equation approach.
Sophisticated techniques aiming at tackling such non-Markovian effects
will be further explored in the future. Other possible future directions
may include generalizing the discussion here to other types of geometry
of JJA, investigating strong-coupling and strong nonlinear effects,
etc.

\section*{Acknowledgement}

Work by JY, EJ, CE, and ANJ was supported by the U.S. Department of
Energy (DOE), Office of Science, Basic Energy Sciences (BES) under
Award \pur{No.~DE-SC0017890.} Work by KLH was funded by ANR BOCA.

\appendix

\section{\label{sec:GKSL}Derivation of the GKSL master equation}

In this section, we start with Eqs.~(\ref{eq:HB-general}),~(\ref{eq:HS})
and~(\ref{eq:HI}) to derive the GKSL master equation, following
the standard procedure in Refs.~{[}\onlinecite{CohenAtomPhoton}{]}
and~{[}\onlinecite{breuer2007thetheory}{]}. Assuming 
\begin{equation}
\rho_{\mathrm{tot}}(t)=\rho_{\mathrm{S}}(t)\otimes\rho_{\mathrm{B}}.
\end{equation}
By moving to the interaction picture associated with the free Hamiltonian
Eqs.~(\ref{eq:HB-general}) and~(\ref{eq:HS}), we obtain the following
Redfield master equation 
\begin{equation}
\frac{\mathrm{d}\rho_{\mathrm{S}}}{\mathrm{d}t}=-\int_{0}^{t}\mathrm{d}t^{\prime}\Tr_{\mathrm{B}}[H_{\mathrm{I}}(t),\,[H_{\mathrm{I}}(t^{\prime}),\,\rho_{\mathrm{S}}(t)\otimes\rho_{\mathrm{B}}]],\label{eq:RedfieldME}
\end{equation}
where
\begin{equation}
H_{\mathrm{I}}(t)\equiv\e^{\ii(H_{\mathrm{S}}+H_{\mathrm{B}})t}H_{\mathrm{I}}\e^{\ii(H_{\mathrm{S}}+H_{\mathrm{B}})t}.
\end{equation}
In Eq.~(\ref{eq:RedfieldME}), we have used the fact that $\Tr_{\mathrm{B}}(H_{\mathrm{I}}\rho_{\mathrm{B}})=0$, which has been justified below Eq.~(\ref{eq:normalization}) in the main text.
Performing the Born--Markov approximation, we obtain 
\begin{equation}
\begin{aligned}[t] & \int_{0}^{t}\mathrm{d}t^{\prime}\Tr_{\mathrm{B}}[H_{\mathrm{I}}(t),\,[H_{\mathrm{I}}(t^{\prime}),\,\rho_{\mathrm{S}}(t)\otimes\rho_{\mathrm{B}}]]\\
 & =\int_{0}^{\infty}\mathrm{d}s\,\Gamma(s)[\mathcal{Q}(t)\mathcal{Q}(t-s)\rho_{\mathrm{S}}(t)-\mathcal{Q}(t)\rho_{\mathrm{S}}(t)\mathcal{Q}(t-s)]+\text{h.c.},
\end{aligned}
\label{eq:double-comm}
\end{equation}
where $\Gamma(s)$ is defined as Eq. (\ref{eq:Gamma-t}) and 
\begin{equation}
\mathcal{Q}(t)\equiv\e^{\ii H_{\mathrm{S}}t}\mathcal{Q}\e^{-\ii H_{\mathrm{S}}t}.
\end{equation}
Therefore, we find 
\begin{equation}
\mathcal{Q}(t)=-\ii\sqrt{\frac{\mathcal{C}\omega_{0}}{2}}(b\e^{-\ii\omega t}-b^{\dagger}\e^{\ii\omega t}).\label{eq:calQ-t}
\end{equation}
Substituting Eq.~(\ref{eq:calQ-t}) into the first term on the right-hand side of Eq.~(\ref{eq:double-comm}), one obtains 
\begin{equation}
\begin{aligned}[t] & \int_{0}^{\infty}\mathrm{d}s\,\Gamma(s)\mathcal{Q}(t)\mathcal{Q}(t-s)\rho_{\mathrm{S}}(t)\\
 & =\frac{\mathcal{C}\omega_{0}}{2}\int_{0}^{\infty}\mathrm{d}s\,\begin{aligned}[t] & \Gamma(s)(b\e^{-\ii\omega_{0}t}-b^{\dagger}\e^{\ii\omega_{0}t})\\
 & \times(b^{\dagger}\e^{\ii\omega_{0}(t-s)}-b\e^{-\ii\omega_{0}(t-s)})\rho_{\mathrm{S}}(t)
\end{aligned}
\\
 & =\frac{\mathcal{C}\omega_{0}}{2}[\Gamma(-\omega_{0})bb^{\dagger}+\Gamma(\omega_{0})b^{\dagger}b]\rho_{\mathrm{S}}(t),
\end{aligned}
\label{eq:int1}
\end{equation}
where we have performed the secular approximation to drop the anti-rotating
and energy nonconservation terms. Similarly, we find 
\begin{multline}
\int_{0}^{\infty}\mathrm{d}s\,\Gamma(s)\mathcal{Q}(t)\rho_{\mathrm{S}}(t)\mathcal{Q}(t-s)\\
=\frac{\mathcal{C}\omega_{0}}{2}[\Gamma(-\omega)b\rho_{\mathrm{S}}(t)b^{\dagger}+\Gamma(\omega)b^{\dagger}\rho_{\mathrm{S}}(t)b].\label{eq:int2}
\end{multline}
Defining $\gamma(\omega_{0})$ and $s(\omega_{0})$ as 
\begin{align}
\gamma(\omega_{0}) & =2\Re\Gamma(\omega_{0}),\label{eq:gamma}\\
s(\omega_{0}) & =-\Im\Gamma(\omega_{0}),\label{eq:s}
\end{align}
we may rewrite 
\begin{align}
 & \Gamma(-\omega_{0})bb^{\dagger}\rho_{\mathrm{S}}(t)+\text{h.c.}=\begin{aligned}[t] & \frac{1}{2}\gamma(-\omega_{0})\{\rho_{\mathrm{S}}(t),\,bb^{\dagger}\}\\
 & +\mathrm{i}s(-\omega_{0})[bb^{\dagger},\,\rho_{\mathrm{S}}(t)],
\end{aligned}
\label{eq:id1}\\
 & \Gamma(\omega_{0})b^{\dagger}b\rho_{\mathrm{S}}(t)+\text{h.c.}=\begin{aligned}[t] & \frac{1}{2}\gamma(\omega_{0})\{\rho_{\mathrm{S}}(t),\,b^{\dagger}b\}\\
 & +\mathrm{i}s(\omega_{0})[b^{\dagger}b,\,\rho_{\mathrm{S}}(t)],
\end{aligned}
\\
 & \Gamma(\omega_{0})b^{\dagger}\rho_{\mathrm{S}}(t)b+\text{h.c.}=\gamma(\omega_{0})b^{\dagger}\rho_{\mathrm{S}}(t)b,\\
 & \Gamma(-\omega_{0})b\rho_{\mathrm{S}}(t)b^{\dagger}+\text{h.c.}=\gamma(-\omega_{0})b\rho_{\mathrm{S}}(t)b^{\dagger}.\label{eq:id4}
\end{align}
Substituting Eqs.~(\ref{eq:double-comm}),~(\ref{eq:int1}),~(\ref{eq:int2}),
~(\ref{eq:id1}) and~(\ref{eq:id4}) into Eq.~(\ref{eq:RedfieldME})
gives 
\begin{equation}
\frac{\mathrm{d}\rho_{\mathrm{S}}}{\mathrm{d}t}=-\mathrm{i}[\rho_{\mathrm{S}}(t),\,H_{\mathrm{LS}}]+\kappa(\omega_{0})\mathcal{D}[b]\rho_{\mathrm{S}}(t)+\cancel{\kappa(-\omega_{0})}\mathcal{D}[b^{\dagger}]\rho_{\mathrm{S}}(t),\label{eq:GKSL-full}
\end{equation}
where 
\begin{align}
 & H_{\mathrm{LS}}=\frac{\mathcal{C}\omega_{0}}{2}\left\{ [s(\omega)+s(-\omega)]b^{\dagger}b+s(-\omega)\right\} ,\\
 & \kappa(\omega_{0})=\frac{\mathcal{C\omega}_{0}}{2}\gamma(\omega_{0}),\\
 & \mathcal{D}[A]\rho_{\mathrm{S}}(t)\equiv A^{\dagger}\rho_{\mathrm{S}}(t)A-\frac{1}{2}\{\rho_{\mathrm{S}}(t),\,A^{\dagger}A\}.
\end{align}
Note that according to Eq.~(\ref{eq:gamma-omega}) in the main text,
the spontaneous absorption rate $\kappa(-\omega_{0})$ is zero. Therefore
Eq.~(\ref{eq:GKSL-full}) reduces to Eq.~(\ref{eq:GKSL}) in the
main text.

\section{\label{sec:Perturbative-evaluation}Perturbative evaluation of the
single-junction two-point correlation function at low temperature}

\subsection{Time-independent perturbation method\label{App_PerturbTheory}}

We consider a single Josephson junction in the regime~$\EJ\ll\EC$.
The corresponding Hamiltonian is $H=H_{\CC}+\lambda H_{\JJ}$, with
$H_{\CC}=\EC N^{2}$, $H_{\JJ}=-\EC\cos\varphi$, and $\lambda=\EJ/\EC\ll1$.
Since the Hamiltonian commutes with the charge conjugation operator~$C$,
we choose to work in a common eigenbasis of $H$ and $C$. The states
of such basis are denoted by $\ket{\psi_{n,\pm}}$, with 
\begin{align}
 & H\ket{\psi_{n,\pm}}=E_{n,\pm}\ket{\psi_{n,\pm}},\label{evals-eq}\\
 & C\ket{\psi_{n,\pm}}=\pm\ket{\psi_{n,\pm}}
\end{align}

The Josephson Hamiltonian~$H_{\JJ}$ will be treated as a perturbation,
$\lambda$ being assumed to be small. Applying, time-independent perturbation
theory\citep{CohenTannoudji2} to lowest nonvanishing order in $\lambda$,
we derive the correlation function~$G(t)=\braket{N(t)N(0)}$, where
time dependence indicates that we consider operators in the Heisenberg
picture (with respect to Hamiltonian~$H$). The average value is
taken over the thermal state with inverse temperature $\beta$,
\begin{equation}
\rho=\frac{1}{Z}\e^{-\beta H},
\end{equation}
where $Z=\Tr\e^{-\beta H}$ is the partition function. The correlation
function is then expanded as follows 
\begin{equation}
\begin{aligned}[t]G(t) & =\frac{\Tr(\e^{-\beta H}\e^{\ii Ht}N\e^{-\ii Ht}N)}{\Tr\e^{-\beta H}}\\
 & =\frac{\sum_{m,n,\pm}\abs{\braket{\psi_{m,\pm}|N|\psi_{n,\mp}}}^{2}\e^{-\beta E_{m,\pm}}\e^{\ii(E_{m,\pm}-E_{n,\mp})t}}{\sum_{n,\pm}\e^{-\beta E_{n,\pm}}}.
\end{aligned}
\label{g}
\end{equation}
In the numerator above, we have taken into account the fact that the
number operator~$N$ only couples states of different charge parities.

Let us now compute the approximate eigenstates and eigenenergies of
the Josephson junction Hamiltonian using time-independent perturbation
theory. We consider the following expansions in powers of $\lambda$,
\begin{align}
 & \ket{\psi_{n,\pm}}=\sum_{q=0}^{\infty}\lambda^{q}\ket{\estate{n,\pm}q},\\
 & E_{n,\pm}=\sum_{q=0}^{\infty}\lambda^{q}\enrg{n,\pm}q.
\end{align}
Eq.~(\ref{evals-eq}) then becomes 
\begin{equation}
\qty(H_{\CC}+\lambda H_{\JJ})\sum_{q=0}^{\infty}\lambda^{q}\ket{\estate{n,\pm}q}=\qty(\sum_{q=0}^{\infty}\lambda^{q}\enrg{n,\pm}q)\sum_{q=0}^{\infty}\lambda^{q}\ket{\estate{n,\pm}q}.
\end{equation}
Sorting out the terms of same order in $\lambda$ in the above equation,
we obtain 
\begin{equation}
H_{\CC}\ket{\estate{n,\pm}q}+H_{\JJ}\ket{\estate{n,\pm}{q-1}}=\sum_{p=0}^{q}\enrg{n,\pm}{p}\ket{\estate{n,\pm}{q-p}}.\label{PerturbExp}
\end{equation}

To zeroth order in $\lambda$, we simply obtain $H_{\CC}\ket{\estate{n,\pm}0}=\enrg{n,\pm}0\ket{\estate{n,\pm}0}$.
This means that $\enrg{n,\pm}0$ is an eigenenergy of $H_{\CC}$,
so we can identify $\enrg{n,\pm}0=n^{2}\EC$. However, the corresponding
eigenstate~$\ket{\estate{n,\pm}0}$ cannot be fully characterized
at this stage. Indeed, all energy levels except the ground state are
two-fold degenerate: the charge states $\ket n$ and $\ket{-n}$ correspond
to the same eigenenergy~$n^{2}\EC$, which means that any linear
combination of these states is also an eigenstate of $H_{\CC}$ with
the same eigenenergy. Consequently, $\ket{\estate{n,\pm}0}$ can be
any such combination. Here, this issue can be resolved invoking charge
parity. Indeed, since the exact eigenstate~$\ket{\psi_{n,\pm}}$
has a definite parity, all the corrections~$\ket{\estate{n,\pm}q}$,
in particular $\ket{\estate{n,\pm}0}$, must too. The only states
of definite parity that can be constructed from the charge states~$\ket n$
and~$\ket{-n}$ are 
\begin{equation}
\ket{\chi_{n,\pm}}=\frac{1}{\sqrt{2}}(\ket n\pm\ket{-n}).
\end{equation}
In conclusion, we choose $\ket{\estate{n,\pm}0}=\ket{\chi_{n,\pm}}$
for the excited states ($n>0$), but we simply have $\ket{\estate00}=\ket0$
for the ground state as it is not degenerate.


To first order in $\lambda$, Eq.~(\ref{PerturbExp}) yields 
\begin{equation}
(H_{\CC}-\enrg{n,\pm}0)\ket{\estate{n,\pm}1}+(H_{\JJ}-\enrg{n,\pm}1)\ket{\estate{n,\pm}0}=0.\label{Exp1}
\end{equation}
We obtain the first-order energy correction by projecting this equation
onto $\ket{\estate{n,\pm}0}$, 
\begin{equation}
\enrg{n,\pm}1=\braket{\estate{n,\pm}0|H_{\JJ}|\estate{n,\pm}0}
\end{equation}
It is clear that $\enrg{n,\pm}1=0$ because $H_{\JJ}$ only couples
neighboring charge states, 
\begin{equation}
\braket{m|H_{\JJ}|n}=-\frac{\EC}{2}(\delta_{m,n+1}+\delta_{m,n-1}).
\end{equation}
As such, the perturbation will not induce any correction to the energy
to first order in $\lambda$. However, there still are corrections
to the states. Indeed, projecting Eq.~(\ref{Exp1}) onto $\ket{\estate{m,\pm}0}$,
$m\ne n$, we find 
\begin{equation}
\braket{\estate{m,\pm}0|\estate{n,\pm}1}=-\frac{\braket{\estate{m,\pm}0|H_{\JJ}|\estate{n,\pm}0}}{\enrg m0-\enrg n0}=\frac{\braket{\estate{m,\pm}0|\!\cos\varphi|\estate{n,\pm}0}}{m^{2}-n^{2}}.
\end{equation}
Note that, for all $m$, $\braket{\estate{m,\mp}0|\estate{n,\pm}1}=0$
since states of different charge parities do not overlap. In particular,
$\braket{\estate{n,\mp}0|\estate{n,\pm}1}=0$. At this stage, only
the component of $\ket{\estate{n,\pm}1}$ along $\ket{\estate{n,\pm}0}$
is still undetermined. It can be obtained invoking the normalization
of the exact eigenstate, $\braket{\psi_{n,\pm}|\psi_{n,\pm}}=1$.
Furthermore, we set the phase of $\ket{\psi_{n,\pm}}$ by imposing
$\braket{\estate{n,\pm}0|\psi_{n,\pm}}\in\mathbb{R}$. To first order
in $\lambda$, this yields $\braket{\estate{n,\pm}0|\estate{n,\pm}1}=0$.
We then conclude 
\begin{equation}
\ket{\estate{n,\pm}1}=\sum_{m\ne n}\frac{\braket{\estate{m,\pm}0|\!\cos\varphi|\estate{n,\pm}0}}{m^{2}-n^{2}}\ket{\estate{m,\pm}0}.
\end{equation}
For $n=0$, this yields 
\begin{equation}
\ket{\estate01}=\frac{1}{\sqrt{2}}\ket{\chi_{1,+}}.
\end{equation}
For $n=1$, we have 
\begin{align}
 & \ket{\estate{1,+}1}=\frac{1}{6}\ket{\chi_{2,+}}+\frac{1}{\sqrt{2}}\ket0,\\
 & \ket{\estate{1,-}1}=\frac{1}{6}\ket{\chi_{2,-}}.
\end{align}
Finally, for $n>1$, we find 
\begin{equation}
\ket{\estate{n,\pm}1}=\frac{1}{2}\qty(\frac{1}{2n+1}\ket{\chi_{n+1,\pm}}-\frac{1}{2n-1}\ket{\chi_{n-1,\pm}}).
\end{equation}

To second order in $\lambda$, Eq.~(\ref{PerturbExp}) yields 
\begin{equation}
(H_{\CC}-\enrg{n,\pm}0)\ket{\estate{n,\pm}2}+(H_{\JJ}-\enrg{n,\pm}1)\ket{\estate{n,\pm}1}=\enrg{n,\pm}2\ket{\estate{n,\pm}0}.
\end{equation}
As before, we project this equation onto $\ket{\estate{n,\pm}0}$
to find 
\begin{equation}
\enrg{n,\pm}2=\braket{\estate{n,\pm}0|H_{\JJ}|\estate{n,\pm}1}=-\EC\sum_{m\ne n}\frac{\abs{\braket{\estate{m,\pm}0|\!\cos\varphi|\estate{n,\pm}0}}^{2}}{m^{2}-n^{2}}.
\end{equation}
For $n=0$, this yields 
\begin{equation}
\enrg02=-\frac{\EC}{2}.
\end{equation}
For $n=1$, we find 
\begin{align}
 & \enrg{1,+}2=\frac{5\EC}{12},\\
 & \enrg{1,-}2=-\frac{\EC}{12}.
\end{align}
Interestingly, the degeneracy of the first excited states is lifted
here. However, this is not the case for higher-energy excited states
since, for $n>1$, we have 
\begin{equation}
\enrg{n,\pm}2=\frac{\EC}{2(4n^{2}-1)}.
\end{equation}

We can now use the results of our perturbative calculation to derive
the correlation function in Eq.~(\ref{g}). For simplicity, corrections
to the energies will be neglected in the oscillating exponentials.
This approximation is justified for short times, $t\ll1/(\lambda^{2}\EC)$.
This is not an issue as the main purpose of our study is to describe
the short-time dynamics of correlations. We can perform the same type
of approximation for the Boltzmann factors, which corresponds to the
high-temperature regime~$\beta^{-1}\gg\lambda^{2}\EC$. More generally,
our perturbative calculation of the energies to second order in $\lambda$
seems insufficient to access the long times or low temperatures, $t\gtrsim1/(\lambda^{3}\EC)$
or $\beta^{-1}\lesssim\lambda^{3}\EC$.

To leading order in $\lambda$, we find that only the states~$\ket{\psi_{n,\pm}}$
and~$\ket{\psi_{n,\mp}}$, degenerate when $\lambda=0$ but corresponding
to different charge parities, contribute to correlations, 
\begin{equation}
\braket{\psi_{m,\pm}|N|\psi_{n,\mp}}=m\delta_{mn}+O(\lambda).
\end{equation}
The correlation function can then be approximated by 
\begin{equation}
G(t)\approx\frac{2\sum_{n=1}^{\infty}n^{2}\e^{-n^{2}\beta\EC}}{1+2\sum_{n=1}^{\infty}\e^{-n^{2}\beta\EC}}.\label{g_highT}
\end{equation}

Actually, it is also possible to obtain the correlation function in
the low-temperature regime. In this case, we neglect the contribution
of excited states to the summations in Eq.~(\ref{g}). In this context,
the Boltzmann factors $\e^{\beta E_{0}}$ cancel each other we do
not need an accurate expression for $E_{0}$. The typical energy gap
between two energy levels is the charging energy~$\EC$---or rather
$\EC$ is a lower bound of the gap---, so this approximation is typically
justified when $\beta^{-1}\ll\EC$. The correlation function is then
approximated by 
\begin{equation}
G(t)\approx\sum_{n}\abs{\braket{\psi_{0}|N|\psi_{n,-}}}^{2}\e^{-\ii(E_{n,-}-E_{0})t}.
\end{equation}
To leading order in $\lambda$, only the state~$\ket{\psi_{1,-}}$
contributes to the summation above, $\braket{\psi_{0}|N|\psi_{n,-}}=\lambda/\sqrt{2}+O(\lambda^{2})$.
As a result, we find, for $t\ll1/(\lambda^{2}\EC)$, 
\begin{equation}
G(t)\approx\frac{\lambda^{2}}{2}\e^{-\ii\EC t}.\label{g_lowT}
\end{equation}

As the temperature is increased, we can include the contributions
from excited states in the summations of Eq.~(\ref{g}) to analyze the transition from Eq.~(\ref{g_lowT}) to Eq.~(\ref{g_highT}).
For example, when $\e^{-\beta\EC}\sim\lambda^{2}$, it becomes relevant
to take into account the contribution of the first excited states
to lowest order in $\lambda$. The correlation function can then be
approximated by 
\begin{equation}
G(t)\approx\frac{\lambda^{2}}{2}\e^{-\ii\EC t}+2\e^{-\beta\EC}.
\end{equation}

\subsection{Matsubara formalism\label{subsec:matusbara}}

The two-point correlation function for a single junction
can \Green{alternatively} be calculated from the Matsubara formalism.\citep{das1997finitetemperature,mahan2013many}
Starting from Eqs.~(\ref{eq:HC-singleJ}) and~(\ref{eq:HJ-singleJ}),
we analytically continue to the imaginary time and obtain the Schrödinger
equation in the interaction picture as 
\begin{equation}
\partial_{\tau}U_{I}(\tau)=\lambda H_{\mathrm{I}}(\tau)U_{I}(\tau),\label{eq:dUdtau}
\end{equation}
where $H_{\mathrm{I}}(\tau)=-\e^{\tau H_{\CC}}H_{\JJ}\e^{-\tau H_{\CC}}$.
The imaginary time propagator $U_{I}(\tau)$ can be explicitly expressed
as 
\begin{equation}
U_{I}(\tau)=\e^{\tau H_{\CC}}\e^{-\tau H},\label{eq:Uoftau}
\end{equation}
which can be verified by substituting into Eq.~(\ref{eq:dUdtau}).
With Eqs.~(\ref{eq:dUdtau}) and~(\ref{eq:Uoftau}), we find 
\begin{align}
 & \Tr[\e^{-\beta H}]=\Tr[\e^{-\beta H_{\CC}}U_{I}(\beta)],\label{eq:Tr1}\\
 & \Tr[\e^{-\beta H}N(\tau)]=\Tr[\e^{-\beta H_{\CC}}U_{I}(\beta)N(0)],
\end{align}
and 
\begin{align}
\Tr[\e^{-\beta H}N(\tau)N(0)] & =\Tr[e^{-\beta H_{\CC}}U_{I}(\beta)U_{I}^{-1}(\tau)N_{I}(\tau)U_{I}(\tau)N(0)]\nonumber \\
 & =\Tr[\e^{-\beta H_{\CC}}\mathcal{T}\left(U_{I}(\beta)N_{I}(\tau)N(0)\right)],\label{eq:Tr3}
\end{align}
where $N_{I}(\tau)\equiv e^{\tau H_{\CC}}N(0)e^{-\tau H_{\CC}}$ is
the number operator in the interaction picture. To the second order
of $\lambda$, we find 
\begin{equation}
\begin{aligned}[t]U_{I}(\tau)= & 1+\lambda\int_{0}^{\tau}\mathrm{d}\tau^{\prime}H_{\mathrm{I}}(\tau^{\prime})\\
 & +\lambda^{2}\int_{0}^{\tau}\mathrm{d}\tau^{\prime}\int_{0}^{\tau^{\prime}}\mathrm{d}\tau^{\prime\prime}H_{\mathrm{I}}(\tau^{\prime})H_{\mathrm{I}}(\tau^{\prime\prime})+\cdots.\label{eq:UI}
\end{aligned}
\end{equation}
In the free rotor basis $\braket{\varphi\big|n}=e^{in\varphi}/\sqrt{2\pi}$,
which are eigenstates of $N$ and $H_{\CC}$ with eigenvalues $n$
and $n^{2}\EC$ respectively, we find 
\begin{align}
 & \braket{n\big|\cos\varphi\big|m}=\frac{1}{2}(\delta_{n+1,\,m}+\delta_{n-1,\,m}),\\
 & \braket{n\big|\cos\varphi\big|k}\braket{k\big|\cos\varphi\big|m}=\frac{1}{4}(\begin{aligned}[t] & \delta_{n+1,\,k}\delta_{k+1,\,m}+\delta_{n+1,\,k}\delta_{k-1,\,m}\\
 & +\delta_{n-1,\,k}\delta_{k+1,\,m}+\delta_{n-1,\,k}\delta_{k-1,\,m}).
\end{aligned}
\end{align}
Therefore, 
\begin{align}
 & \braket{n\big|H_{\mathrm{I}}(\tau^{\prime})\big|n}=0,\label{eq:HI-DME}\\
 & \braket{n\big|H_{\mathrm{I}}(\tau^{\prime})\big|k}\braket{k\big|H_{\mathrm{I}}(\tau^{\prime\prime})\big|n}=\frac{\EC^{2}}{4}[\begin{aligned}[t] & \delta_{n+1,\,k}F_{n}(\tau^{\prime}-\tau^{\prime\prime})\\
 & +\delta_{n-1,\,k}F_{-n}(\tau^{\prime}-\tau^{\prime\prime})],
\end{aligned}
\label{eq:HIHI-DME}
\end{align}
where 
\begin{equation}
F_{n}(\tau)=\exp\left[-\EC\tau(1+2n)\right].\label{eq:Fn}
\end{equation}
Eq. (\ref{eq:Tr1}) can be rewritten as 
\begin{equation}
\Tr[\e^{-\beta H}]=\sum_{n}\e^{-\beta\EC n^{2}}\braket{n\big|U_{I}(\beta)\big|n}.\label{eq:partition}
\end{equation}
To the second order of $\lambda$, according to Eqs. (\ref{eq:UI},
\ref{eq:HI-DME}, \ref{eq:HIHI-DME}), we find 
\begin{equation}
\braket{n\big|U_{I}(\beta)\big|n}=1-\frac{\lambda^{2}\EC^{2}}{4}[K{}_{n}(\beta)+K_{-n}(\beta)]+O(\lambda^{3}),\label{eq:UI-DME}
\end{equation}
where 
\begin{equation}
\begin{aligned}[t]K_{n}(\tau) & =\int_{0}^{\tau}\mathrm{d}\tau^{\prime}\int_{0}^{\tau^{\prime}}\mathrm{d}\tau^{\prime\prime}F_{n}(\tau^{\prime}-\tau^{\prime\prime})\\
 & =\begin{aligned}[t] & \frac{1}{(1+2n)^{2}\EC^{2}}\e^{-\tau\EC(1+2n)}\\
 & +\frac{\tau}{(1+2n)\EC}-\frac{1}{(1+2n)^{2}\EC^{2}}.
\end{aligned}
\end{aligned}
\label{eq:Kn}
\end{equation}
The infinite series in the right hand side of Eq. (\ref{eq:partition})
can be evaluated at the limit $\beta\EC\gg1$, where the infinite
sum is replaced by the lowest order in $\e^{-\beta\EC}$ in the summand.
Therefore 
\begin{equation}
\begin{aligned}[t]\sum_{n}\e^{-\beta\EC n^{2}}K_{n}(\beta) & =\sum_{n}\e^{-\beta\EC n^{2}}K_{-n}(\beta)\\
 & =\begin{aligned}[t] & \sum_{n}\frac{1}{(1+2n)^{2}\EC^{2}}\e^{-\beta\EC(n+1)^{2}}\\
 & +\sum_{n}\left[\frac{\beta}{(1+2n)\EC}-\frac{1}{(1+2n)^{2}\EC^{2}}\right]\e^{-\beta\EC n^{2}}
\end{aligned}
\\
 & =\frac{\beta}{\EC}.
\end{aligned}
\end{equation}
Therefore we find 
\begin{equation}
\Tr[\e^{-\beta H}]=1-\frac{\beta\EC\lambda^{2}}{2}+O(\lambda^{2})O(e^{-\beta\EC}).
\end{equation}
According to Eq.~(\ref{eq:UI-DME}), we observe that $\braket{n\big|U_{I}(\beta)\big|n}$
is even in $n$ up the second order in $\lambda$. Therefore, we find
\begin{equation}
\Tr[\e^{-\beta H_{\CC}}U_{I}(\beta)N(0)]=\sum_{n}n\e^{-n^{2}\beta\EC}\braket{n\big|U_{I}(\beta)\big|n}=O(\lambda^{3})
\end{equation}
to all orders of $e^{-\beta\EC}$. To evaluate Eq. (\ref{eq:Tr3}),
let us first calculate the following 
\begin{equation}
\begin{aligned}[t] & \braket{n\big|\mathcal{T}\left(U_{I}(\beta)N_{I}(\tau)N(0)\right)\big|n}\\
 & =\begin{aligned}[t] & \braket{n\big|N_{I}(\tau)N(0)\big|n}\\
 & +\lambda\cancel{\int_{0}^{\beta}\mathrm{d}\tau^{\prime}\braket{n\big|\mathcal{T}\left(H_{\mathrm{I}}(\tau^{\prime})N_{I}(\tau)Q(0)\right)\big|n}}\\
 & +\frac{\lambda^{2}}{2}\int_{0}^{\beta}\mathrm{d}\tau^{\prime}\int_{0}^{\beta}\mathrm{d}\tau^{\prime\prime}\braket{n\big|\mathcal{T}\left(H_{\mathrm{I}}(\tau^{\prime})H_{\mathrm{I}}(\tau^{\prime\prime})N_{I}(\tau)Q(0)\right)\big|n}\\
 & +O(\lambda^{3}),
\end{aligned}
\end{aligned}
\label{eq:TUNN}
\end{equation}
where the second term in the right hand side vanish is due to Eq.~(\ref{eq:HI-DME}).
Now let us evaluate the last term in the right hand side of Eq.~(\ref{eq:TUNN}).
It can be written as four parts $\lambda^{2}/2\sum_{k=1}^{4}I_{kn}(\tau)$,
where 
\begin{align}
 & I_{1n}(\tau)=\int_{0}^{\tau}\mathrm{d}\tau^{\prime}\int_{0}^{\tau}\mathrm{d}\tau^{\prime\prime}\braket{n\big|N_{I}(\tau)\mathcal{T}\left(H_{\mathrm{I}}(\tau^{\prime})H_{\mathrm{I}}(\tau^{\prime\prime})\right)N(0)\big|n},\\
 & I_{2n}(\tau)=\int_{0}^{\tau}\mathrm{d}\tau^{\prime}\int_{\tau}^{\beta}\mathrm{d}\tau^{\prime\prime}\braket{n\big|H_{\mathrm{I}}(\tau^{\prime\prime})N_{I}(\tau)H_{\mathrm{I}}(\tau^{\prime})N(0)\big|n},\\
 & I_{3n}(\tau)=\int_{\tau}^{\beta}\mathrm{d}\tau^{\prime}\int_{0}^{\tau}\mathrm{d}\tau^{\prime\prime}\braket{n\big|H_{\mathrm{I}}(\tau^{\prime})N_{I}(\tau)H_{\mathrm{I}}(\tau^{\prime\prime})N(0)\big|n},\\
 & I_{4n}(\tau)=\int_{\tau}^{\beta}\mathrm{d}\tau^{\prime}\int_{\tau}^{\beta}\mathrm{d}\tau^{\prime\prime}\braket{n\big|\mathcal{T}\left(H_{\mathrm{I}}(\tau^{\prime})H_{\mathrm{I}}(\tau^{\prime\prime})\right)N_{I}(\tau)N(0)\big|n},
\end{align}

Using the short hand notation introduced in Eqs.~(\ref{eq:Fn}) and~(\ref{eq:4th-integral}),
we find 
\begin{align}
 & I_{1n}(\tau)=\frac{\EC^{2}}{2}[n^{2}K_{n}(\tau)+n^{2}K_{-n}(\tau)],\\
 & I_{2n}(\tau)=\frac{\EC^{2}}{4}[n(n+1)L_{n}(\tau)+n(n-1)L_{-n}(\tau)],\\
 & I_{3n}(\tau)=I_{2n}(\tau),\\
 & I_{4n}(\tau)=\frac{\EC^{2}}{2}[n^{2}K_{n}(\beta-\tau)+n^{2}K_{-n}(\beta-\tau)],
\end{align}
where 
\begin{align}
 & L_{n}(\tau)\nonumber \\
= & \int_{0}^{\tau}\mathrm{d}\tau^{\prime}\int_{\tau}^{\beta}\mathrm{d}\tau^{\prime\prime}F_{n}(\tau^{\prime\prime}-\tau^{\prime})=\int_{\tau}^{\beta}\mathrm{d}\tau^{\prime}\int_{0}^{\tau}\mathrm{d}\tau^{\prime\prime}F_{n}(\tau^{\prime}-\tau^{\prime\prime})\nonumber \\
= & \frac{1}{(1+2n)^{2}\EC^{2}}\bigg\{\e^{-\beta\EC(1+2n)}\left(1-\e^{\tau\EC(1+2n)}\right)-\e^{-\tau\EC(1+2n)}+1\bigg\},\label{eq:Ln}
\end{align}
and we have used the fact that 
\begin{equation}
\begin{aligned}[t]\int_{\tau}^{\beta}\mathrm{d}\tau^{\prime}\int_{\tau}^{\tau^{\prime}}\mathrm{d}\tau^{\prime\prime}F_{n}(\tau^{\prime}-\tau^{\prime\prime}) & =\int_{0}^{\beta-\tau}\mathrm{d}\tau^{\prime}\int_{0}^{\tau^{\prime}}\mathrm{d}\tau^{\prime\prime}F_{n}(\tau^{\prime}-\tau^{\prime\prime})\\
 & =K_{n}(\beta-\tau).
\end{aligned}
\label{eq:4th-integral}
\end{equation}
The real-time correlation function is 
\begin{equation}
G(t)=\frac{1}{\Tr(\e^{-\beta H})}\sum_{n}\e^{-\beta\EC n^{2}}\left(n^{2}+\frac{\lambda_{\mathrm{}}^{2}}{2}\sum_{k=1}^{4}I_{kn}(\ii t)\right).\label{eq:I4}
\end{equation}
Let us go back to the real time by replacing $\tau\to it$ in Eq.~(\ref{eq:I4})
and then perform the low temperature approximation $e^{-\beta\EC}\ll1$.
We need to evaluate 
\begin{align}
 & \sum_{n}n^{2}K_{n}(\ii t)\e^{-n^{2}\beta\EC}=\sum_{n}n^{2}K_{-n}(\ii t)\e^{-n^{2}\beta\EC}=O(\e^{-\beta\EC}),\\
 & \sum_{n}n(n+1)L_{n}(\ii t)\e^{-\beta\EC n^{2}}=O(e^{-\beta\EC}),\\
 & \sum_{n}n(n-1)L_{-n}(\ii t)\e^{-\beta\EC n^{2}}=O(e^{-\beta\EC}),
\end{align}
\begin{align}
\sum_{n}n^{2}K_{n}(\beta-\ii t)\e^{-\beta\EC n^{2}} & =\sum_{n}n^{2}K_{-n}(\beta-\ii t)\e^{-\beta\EC n^{2}}\nonumber \\
 & =\frac{1}{\EC^{2}}\e^{-\ii\EC t}+O(e^{-\beta\EC}).
\end{align}
Based on these results, one readily observe that the leading order
contributions to the Green's function are of the order $O(e^{-\beta\EC})O(\lambda^{0})$
and $O([e^{-\beta\EC}]^{0})O(\lambda^{2})$. There first contribution
comes from the first term in Eq.~(\ref{eq:I4}) while the second
contribution comes from $I_{4n}(\ii t)$. Therefore we find 
\begin{equation}
G(t)=\frac{2e^{-\beta\EC}+\lambda^{2}\e^{-\ii\EC t}/2+O(\lambda^{2})O(e^{-\beta\EC})}{1-\beta\EC\lambda^{2}/2+O(\lambda^{2})O(e^{-\beta\EC})}.\label{eq:G}
\end{equation}
From the numerator of Eq.~(\ref{eq:G}), one concludes that as long
as $e^{-\beta\EC},\,\lambda\ll1$, the numerator is a good approximation
to $\Tr[N_{I}(t)N(0)\e^{-\beta H}]$. From the denominator of see
that the Matsubara perturbative approach works well if $\beta\EC\lambda^{2}\ll1$.
Keeping only second order of $\lambda^{2}$ and the first order in
$e^{-\beta\EC}$, one can replace the denominator in Eq.~(\ref{eq:G})
with $1$ and obtain Eq.~(\ref{G_eps}). Furthermore, in the zero-temperature
limit given by Eq.~(\ref{eq:LowT}), $e^{-\beta\EC}\ll\lambda^{2}$
and therefore Eq.~(\ref{eq:G}) reduces to Eq.~(\ref{eq:Gi}) in
the main text.

 \bibliographystyle{apsrev4-1}
\bibliography{JJ}

\end{document}